\documentclass[a4paper,11pt]{article}
\usepackage{jheppub} 
\usepackage{lineno}
\usepackage{subcaption}
\usepackage{array}
\usepackage{float}
\usepackage{xcolor}
\usepackage{tikz}
\usetikzlibrary{arrows.meta,positioning,shapes.geometric,calc}

\title{Metric Reconstruction from Timelike Entanglement Entropy}

\author[a,b]{Hao Feng,}
\author[a,c]{Tian-Shun Chen,}
\author[a,d]{and Shao-Feng Wu}

\affiliation[a]{Department of Physics, Shanghai University, Shanghai, 200444, China}
\affiliation[b]{Kavli Institute for Theoretical Sciences (KITS) and School of Quantum,
University of the Chinese Academy of Sciences,
Beijing, 100190, China}
\affiliation[c]{Tsung-Dao Lee Institute, Shanghai Jiao-Tong University, Shanghai 201210, China}
\affiliation[d]{Center for Gravitation and Cosmology, Yangzhou University, Yangzhou 225009, China}

\emailAdd{fenghao26@mails.ucas.ac.cn}
\emailAdd{DeltaChen@sjtu.edu.cn}
\emailAdd{sfwu@shu.edu.cn}

\abstract{Timelike entanglement entropy (TEE) provides a Lorentzian boundary
probe of bulk geometry, but its use for metric reconstruction depends on the
holographic prescription and on the extremal-surface branch selected by that
prescription.  We study this inverse problem for strip-shaped TEE data and make
these dependencies explicit. In the complex-valued weak extremal surface
(CWES) prescription, the time-width dependence of TEE determines an Abel density
\(H(W)\) on a selected real branch; for
Bañados–Teitelboim–Zanelli (BTZ) black holes this gives an analytic
reconstruction of the blackening factor once the singularity endpoint fixes the radial origin. After developing a forward numerical method for the complex-coordinate prescription, we
formulate it as the main reconstruction scheme for the asymptotically
AdS\(_{d+1}\) backgrounds with \(d\ge2\).
On a chosen complex branch, the time-width dependence of TEE supplies the Abel
input that determines the TEE-accessible density, while one additional geometric anchor is required to
convert that density into a definite radial metric profile.  With UV or
horizon-scale anchoring and rational continuation from the reconstructed
complex-path samples, the method reproduces the benchmark BTZ,
four-dimensional Schwarzschild, and Reissner--Nordstr\"om (RN)
blackening factors.  However, the
Gubser--Rocha example shows that a single strip observable with a nontrivial
spatial warp factor determines only one functional combination of the metric
functions.}
\begin{document}
\maketitle
\flushbottom

\section{Introduction}

The anti-de Sitter/conformal field theory (AdS/CFT) correspondence provides a
controlled holographic framework in which spacetime geometry is encoded in
boundary quantum degrees of freedom
\cite{Maldacena_1999,Gubser_1998,witten1998antisitterspaceholography}.  Its
sharp quantum-information entry is holographic entanglement entropy (HEE):
the Ryu--Takayanagi and Hubeny--Rangamani--Takayanagi prescriptions relate
boundary entanglement on spacelike regions to areas of bulk extremal surfaces
\cite{Ryu_2006_1,Ryu_2006_2,Hubeny_2007}.  This relation made entanglement not
only a diagnostic of geometry but also a source of inverse data.  In suitable
settings, boundary entropies, modular data, and complexity-type observables can
be used to reconstruct metric functions, causal wedges, and even the black-hole
interiors
\cite{Hammersley:2006cp,Hammersley:2007ab,Bilson:2008ab,Bilson_2011,Hubeny:2012ry,Balasubramanian:2013lsa,Myers:2014jia,Czech:2014ppa,Engelhardt:2016wgb,Engelhardt:2016crc,Roy:2018ehv,Bao:2019bib,Burda:2018rpb,Hernandez-Cuenca:2020ppu,Jokela:2020auu,Hashimoto_2021,Xu_2023,Chae:2026covariant,Jokela_2023,Ahn_2025,Ji:2025vks}.

Timelike entanglement entropy (TEE) asks whether an analogous information
probe exists for the time direction. The modern formulation grew out of two
closely related ideas.\footnote{Earlier discussions related to timelike entanglement can be found, e.g., in Refs.~\cite{Olson:2011timelike,Wang:2018jva}.}
First, pseudo-entropy extends the von Neumann entropy
from density matrices to reduced transition matrices and is therefore allowed
to be complex \cite{Nakata_2021}.  Second, analytic continuation across the
light cone suggests that the CFT entropy of a spacelike interval can be
continued to a timelike interval.  In two-dimensional CFTs this continuation
produces the familiar logarithmic real part together with an imaginary
constant.  Doi et al. promoted this structure to TEE and interpreted it as a
pseudo-entropy of timelike subregions, with
holographic realizations in AdS$_3$, BTZ, and shock-wave
geometries, as well as connections to dS/CFT
\cite{Doi_2023P,Doi_2023T}.  These works, together with a later
quantum-information perspective, motivate the proposal that pseudo-entropy and
TEE may probe the emergence of a holographic time direction
\cite{Doi_2023P,Doi_2023T,takayanagi2025emergentholographicspacetimequantum}.

Much of the subsequent work can be organized along two lines.  One is the
extension of TEE to diverse theories and holographic settings.
Deformations, nonconformal theories, and renormalization-group flows have
been
explored~\cite{jiang2024timelikeentanglemententropytbart,afrasiar2024timelikeentanglemententropyphase,Grieninger_2024,giataganas2025timelikeentanglementrg},
alongside dS and AdS/BCFT
backgrounds~\cite{Narayan_2023,jiang2024timelikeentanglemententropyds3cft2,Chu_2023,Narayan:2026dsc}.
Growing interests focus on stationary black holes, probing their
interiors and
singularities~\cite{anegawa2024blackholesingularitytimelike,wen2025timelikegravitationalanomalousentanglement,GuoXu:2025rtduality,guo2025measuringblackholeinterior,Roychowdhury:2025qbtz,afrasiar2025aspectsholographictimelike,Prihadi:2026nua,LassoAndino:2026new,Dai:2026tmp,li2026lineargrowthholographictimelike,Pal:2026htesc}.
Meanwhile, dynamical black holes and Hawking radiation have also been
studied~\cite{Katoch:2025bnh,Ladghami:2026hby,katoch2026entanglementinequalitiestimelikeintervals,Das:2026dte}.
Other extensions cover topics such as non-relativistic
geometries~\cite{jena2024noteholographictimelikeentanglement,afrasiar2024holographictimelikeentanglemententropy},
higher-curvature gravity~\cite{Zhao:2025zgm}, and top-down
constructions~\cite{Roychowdhury_2025,nunez2025timelikeentanglementtopdown,nunez2025holographictimelikeentanglementdimensions}.

Another line addresses the characteristic features, formulations, and
theoretical connections of TEE, including the origin and probe dependence
of its imaginary
part~\cite{guo2024imaginarytimelikeentanglemententropy,chu2025timelikegravitationalanomalies,AliAkbari:2026pdi},
temporal entanglement in quantum systems and field
theory~\cite{Liu_2024,narayan2024notestimeentanglementpseudoentropy,Das:2026sdk,CastroAlvaredo:2026teqft},
its relations to reflected entropy and spacelike
entanglement~\cite{basu2024reflectedentropytimelikeentanglement,guo2024relationtimelikespacelikeentanglement,gong2025entanglementmeasurescausally},
and the interplay between TEE first laws and bulk field
equations~\cite{Li:2025tud,Xiao:2026tfl}.

Despite these developments, there are inequivalent notions
of timelike entanglement and competing prescriptions for their holographic
realization. Analytic-continuation and pseudo-entropy constructions generally yield
complex-valued quantities \cite{Doi_2023P,Doi_2023T}.
Ref. \cite{milekhin2025observablecomputableentanglementtime} instead develops a
microscopic, measurable construction for time-separated subsystems based on a
spacetime density matrix.  In relativistic quantum field theory, the moments of
this matrix agree with analytic continuation, but the resulting entanglement
measure coincides with single-interval timelike pseudo-entropy only in special
cases.  Related work embeds TEE in a broader non-Hermitian density-matrix
framework and connects
non-Hermiticity to causal influence in unitary settings
\cite{harper2025nonhermitiandensitymatrices}.  By contrast, an
operator-algebraic proposal for a single timelike
interval gives a real-valued entropy via the timelike-tube theorem, assuming
additivity, the split property, and primitive causality
\cite{jiang2025timelikeentanglemententropyrevisited}.  At the bulk level, even
after a particular boundary observable has been fixed, a generally applicable
holographic prescription has not been established, and several inequivalent
bulk constructions have been proposed.

The candidate bulk constructions relevant to the present inverse problem fall
into two broad geometric classes.  The first employs mixed-signature surfaces
in a real Lorentzian geometry.  Doi et al.\ define holographic timelike
entanglement entropy (HTEE) from a stationary combination of spacelike and
timelike extremal segments homologous to the timelike boundary region
\cite{Doi_2023P,Doi_2023T}.  The CWES prescription systematizes this idea by
requiring segmentwise extremality, weak extremality under joint variations, and
an ordering relation for complex-valued areas \cite{Li_2023}. A related bulk extremization procedure determining composite
timelike-spacelike geodesics is proposed in Ref.~\cite{Bohra:2025mhb}. In addition, the surface-merging prescription that extends to non-conformal
theories is developed in Ref.~\cite{afrasiar2024timelikeentanglemententropyphase}. The second class
treats HTEE as the area of an extremal surface in a complexified bulk, whose
real and imaginary parts are generally not associated with separate real
segments 
\cite{Heller_2025,heller2025temporalentanglementholographicentanglement,
nunez2025interpolatingspaceliketimelikeentanglement}.

These constructions can yield the same regulated entropy in special geodesic
settings, but this agreement is not general.  More recently, a boundary
construction based on the time-ordered replica twist correlator was shown to
select complex geodesics in AdS$_3$/CFT$_2$.  In the dynamical
AdS$_3$-Vaidya background, the complex-geodesic result reproduces the explicit
CFT correlator, whereas earlier piecewise-curve constructions yield a different
entropy \cite{bernamonti2026temporalentanglementtwistcorrelators}.  This result
does not directly test CWES or establish a prescription-level
disagreement for the static planar black-hole strip configurations studied
below.  It nevertheless
demonstrates that distinct bulk prescriptions cannot in general be treated as
interchangeable realizations of a fixed boundary TEE.  Metric reconstruction
must therefore keep the boundary input, bulk prescription, and branch
choice explicit.

This paper addresses the inverse problem for strip-shaped holographic TEE
data.  Previous work used timelike modular-Hamiltonian constraints to formulate
an HKLL-type reconstruction of free bulk scalar fields in (A)dS
\cite{Das_2024}.  That operator-reconstruction problem is distinct from the
metric reconstruction from an entropy curve studied here.
Given the time-width dependence on a selected timelike
extremal-surface branch, we ask what bulk geometric information is encoded in
the boundary observable.  Our strategy is to adapt the Abel inversion familiar
from spacelike HEE.  In the spacelike strip problem, the turning point \(z_s\)
is a real radial coordinate, and the relation between the boundary width
\(l(z_s)\) and the exterior blackening factor can be recast as an Abel integral
equation on a real branch
\cite{Bilson_2011,Jokela:2020auu,Xu_2023}.  For timelike regions, the same
square-root integral structure survives, but its geometric meaning changes.  The
transformed variable \(W\) is generally a
branch-dependent combination of metric functions appearing under the
square-root of the extremal-surface integral.  It may be complex-valued and need
not define a global radial coordinate.  The reconstruction therefore yields the
Abel density selected by a chosen prescription and branch.  Converting this
density into a radial metric profile requires a separate geometric anchor.

We first apply the reconstruction to CWES data in
Sec.~\ref{sec:2}. Note that we select the CWES as the representative of the first class of HTEE prescriptions.  We take as input the mixed-signature CWES branch described below for each boundary time width; this branch is fixed upstream of the reconstruction. The Hamilton--Jacobi relation between the conserved
quantity and the derivative of the timelike entropy follows from the area
functional of these surfaces.  The inverse problem then becomes an analytically
continued square-root transform whose discontinuity is a standard Abel
integral equation.  In the BTZ benchmark its inversion reconstructs the
expected metric once the singularity endpoint fixes the radial origin.

In Sec.~\ref{sec:3}, we turn to the complex-coordinate prescription,
which provides the main constructive scheme of this work.  We first develop a
numerical method to compute its HTEE and then formulate the corresponding
inverse reconstruction for
\((d+1)\)-dimensional asymptotically AdS backgrounds and
take a regularized entropy curve \(S_{\rm reg}(\Delta t)\) on a selected
branch as the boundary input.  The
Hamilton--Jacobi relation determines \(E(\Delta t)\) and its
inverse generates the time-width function \(\Delta t(E)\) entering
the Abel equation.  The ordered curve \(W_t=-E^2\) supplies the 
integration endpoint, while tracing complex turning points and contours
specifies the branch on which the data are interpreted.
Abel inversion gives the TEE-accessible density, while a UV
or horizon-scale anchor fixes the remaining radial-profile freedom.  We test
the method in BTZ, four-dimensional Schwarzschild, and Reissner--Nordstr\"om (RN) backgrounds,
where the anchored procedure reproduces the expected blackening functions.  We
also analyze the Gubser--Rocha model~\cite{Gubser_2010}; there a nontrivial
spatial warp factor implies that one strip observable determines only one
functional combination of \(f(z)\) and \(h(z)\), so additional data are needed
to separate the two metric functions.

Finally, Sec.~\ref{sec:conclusion} gives the conclusions and discusses the
scope and limitations of our reconstruction.

\section{Reconstruction from CWES data}
\label{sec:2}
\subsection{CWES input and scope}

For the conditional inverse problem considered in this section, HTEE denotes
the output of a specified bulk extremal-surface prescription for a timelike
boundary subregion.
Once the relevant surface or contour has been specified, the entropy takes the
area form
\begin{equation}
  S_T = \frac{A}{4G_N}.  \label{eq:RT_formula}
\end{equation}
Hereafter, we set the Newton constant and the AdS radius to unity.  Areas and
entropies of strip regions are quoted per unit volume in the \(d-2\) transverse
directions.
The choice of surface or contour is not unique.  Several inequivalent
constructions have been proposed for timelike boundary subregions.  Here we condition the
inverse problem on a selected mixed-signature CWES branch for each boundary time
width.  We assume that the CWES joint-variation and area-ordering conditions
\cite{Li_2023} have already selected this branch and that it varies smoothly
with the width over the interval considered.

\begin{figure}[htbp]
    \centering

    \includegraphics[width=0.5\textwidth]{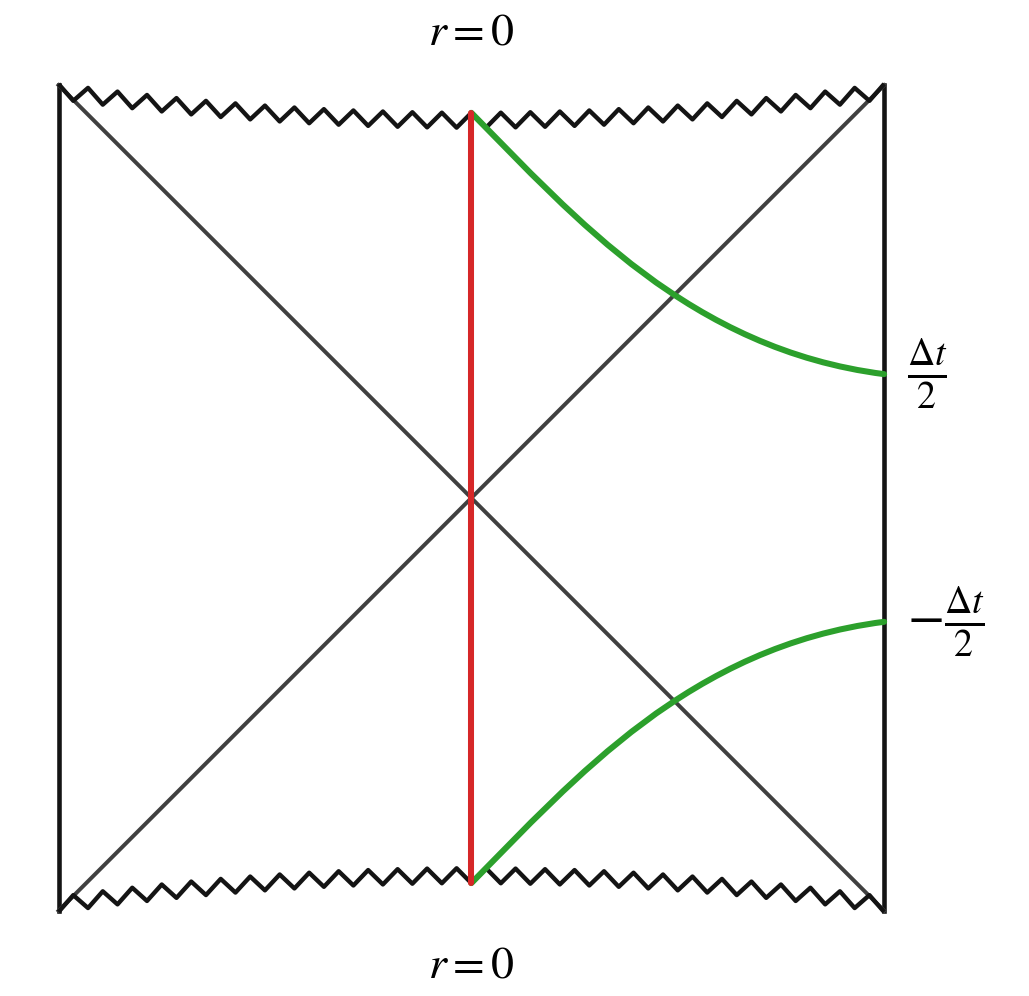}
    \caption{Schematic Penrose diagram illustrating the selected CWES
    branch assumed in the reconstruction.  The timelike boundary interval extends from
    \(-\Delta t/2\) to \(\Delta t/2\).  The two spacelike branches (green)
    connect the boundary endpoints to the future and past singularities, while
    the timelike codimension-two surface (red) connects the two singularity
    joints through the bifurcation surface.}
    \label{fig:penrose_diagram}
\end{figure}

We work with a \((d+1)\)-dimensional bulk spacetime dual to a
\(d\)-dimensional boundary theory with \(d\ge2\), and fix a
spatial direction \(y=0\) to
define a timelike strip-shaped boundary subregion, extended along the $d-2$ transverse spatial directions.  For each boundary time width, the
selected mixed-signature branch shown in
Fig.~\ref{fig:penrose_diagram} \cite{Li:2026fcr,li2026lineargrowthholographictimelike}
consists of two spacelike branches and a timelike codimension-two surface
extended along the transverse directions.  The spacelike branches connect the
two boundary endpoints to the future and past spacelike singularities,
respectively, while the timelike surface connects the two singularity joints
through the bifurcation surface and reduces to a timelike geodesic for \(d=2\).

Suppose that the bulk geometry is described by the static metric
\begin{equation}
ds^{2}=-f(r)\,dt^{2}+\frac{dr^{2}}{f(r)}+r^{2}\left(d\vec{x}^{\,2}+dy^{2}\right).
\end{equation}
Here \(r=0\) denotes a spacelike singular endpoint and \(r=r_h\) denotes the event horizon. We assume that no additional horizon is encountered along the selected CWES branch \cite{li2026lineargrowthholographictimelike}.

We first evaluate the two spacelike branches, which supply the
\(E\)-dependent real part of the TEE. Parametrizing the spacelike branch by \(r=r(t)\) and setting \(dy=0\), the induced metric becomes
\begin{equation}
ds^{2}=\left(-f(r)+\frac{\dot r^{\,2}}{f(r)}\right)dt^{2}+r^{2}d\vec{x}^{\,2},
\end{equation}
where the dot denotes differentiation with respect to \(t\).
The corresponding total area functional is
\begin{equation}
A_{\mathrm{sp}}=2\int dt\,\mathcal{L},
\qquad
\mathcal{L}(r,\dot r)
\equiv r^{d-2}\sqrt{-f(r)+\frac{\dot r^{\,2}}{f(r)}},
\end{equation}
where the factor of \(2\) accounts for the two symmetric spacelike branches.

Since \(\mathcal{L}\) does not depend explicitly on \(t\), there is a conserved quantity
\begin{equation}
E
=
\mathcal{L}
-\dot r\,\frac{\partial\mathcal{L}}{\partial\dot r}
=
-\frac{r^{d-2}f(r)}
{\sqrt{-f(r)+\frac{\dot r^{\,2}}{f(r)}}}.
\end{equation}
Solving for $\dot r$ gives
\begin{equation}
\left(\frac{dt}{dr}\right)^2
=
\frac{E^2}
{f(r)^2\left[r^{2d-4}f(r)+E^2\right]}.
\end{equation}
This equation determines \(dt/dr\) only up to a local sign.  We choose the
branch that runs from \(t=0\) at \(r=0\) to \(t=\Delta t/2\) at the
asymptotic boundary.  The second spacelike branch is obtained by time
reflection. The full boundary temporal width is therefore twice the contribution of
either branch:
\begin{equation}
\Delta t=2E\int_{0}^{+\infty}\frac{dr}{f(r)\sqrt{r^{2d-4}f(r)+E^{2}}}.\label{eq:T0_general}
\end{equation}
Substituting $dt=(dt/dr)\,dr$ back into the area functional, one finds
\begin{equation}
A_{\mathrm{sp}}=2\int_{0}^{+\infty}\frac{r^{2d-4}}{\sqrt{r^{2d-4}f(r)+E^{2}}}\,dr.
\end{equation}
Regulating the standard area-law divergence at \(r=\Lambda\), the real part of the
TEE is
\begin{equation}
\mathrm{Re}\,S_T
=\frac{1}{2}\int_{0}^{\Lambda}\frac{r^{2d-4}}{\sqrt{r^{2d-4}f(r)+E^{2}}}\,dr.\label{eq:ReSA_general}
\end{equation}

The remaining contribution comes from the timelike branch at \(t=0\) inside the
horizon. Its induced area is
\begin{equation}
A_{\mathrm{tm}}=2\int_{0}^{r_{h}}\frac{r^{d-2}}{\sqrt{-f(r)}}\,dr.
\end{equation}
After including the factor \(1/4\), the timelike segment contributes
the following imaginary part to the TEE:
\begin{equation}
\mathrm{Im}\,S_T
=\frac{1}{2}\int_{0}^{r_{h}}\frac{r^{d-2}}{\sqrt{-f(r)}}\,dr.
\end{equation}

Combining the two pieces, the full TEE is
\begin{equation}
S_T
=\frac{1}{2}\int_{0}^{\Lambda}
\frac{r^{2d-4}}{\sqrt{r^{2d-4}f(r)+E^{2}}}\,dr
+i\,\frac{1}{2}\int_{0}^{r_{h}}\frac{r^{d-2}}{\sqrt{-f(r)}}\,dr.\label{eq:TEE_total_general}
\end{equation}

\subsection{Reconstruction framework}

\label{sec:reconstruction_framework}

We now use the area of the assumed CWES as inverse data.  The timelike segment in
Eq.~\eqref{eq:TEE_total_general} contributes an \(E\)-independent imaginary
constant, so it drops out of the variation with respect to
the branch parameter \(E\).  The \(E\)-dependent input is therefore the
spacelike contribution, together with the time-width relation
\eqref{eq:T0_general}.

Differentiating the spacelike entropy contribution
\eqref{eq:ReSA_general} with respect to \(E\) gives the finite result
\begin{equation}
\frac{dS_T}{dE}
=
\frac{d\,\mathrm{Re}\,S_T}{dE}
=
-\frac{E}{2}
\int_{0}^{+\infty}
\frac{r^{2d-4}}{\left(r^{2d-4}f(r)+E^{2}\right)^{3/2}}\,dr.
\label{eq:dSRE_dE_general}
\end{equation}
Similarly, differentiating Eq.~\eqref{eq:T0_general} with respect to \(E\)
gives
\begin{equation}
\frac{d\Delta t}{dE}
=
2\int_{0}^{+\infty}
\frac{r^{2d-4}}{\left(r^{2d-4}f(r)+E^{2}\right)^{3/2}}\,dr.
\label{eq:dT0_dE_general}
\end{equation}
Taking the ratio of these two derivatives, we obtain
\begin{equation}
\frac{dS_T}{d\Delta t}
=
\frac{d\,\mathrm{Re}\,S_T}{d\Delta t}
=
\frac{dS_T/dE}{d\Delta t/dE}
=
-\frac{E}{4}.
\label{eq:HJ_general_dimension}
\end{equation}
This is the Hamilton--Jacobi relation for the spacelike branches of the
assumed CWES.  The same
relation was obtained in Ref.~\cite{Li:2026fcr} from the Hamiltonian
formalism; here it follows directly by differentiating their area, in
a form adapted to the integral-equation reconstruction strategy of
Refs.~\cite{Bilson_2011,Jokela:2020auu,Xu_2023}.  It determines the conserved quantity \(E\)
from the slope of the boundary TEE:
\begin{equation}
E(\Delta t)=-4\frac{dS_T}{d\Delta t}.
\label{eq:E_from_entropy_general}
\end{equation}
This derivation is valid on an interval where the CWES fixed upstream
varies smoothly with \(E\).

Note that Eq.~\eqref{eq:dT0_dE_general} implies
\(\frac{d\Delta t}{dE}>0\) on the branch considered. Hence
\(\Delta t(E)\) is strictly monotonic and can be inverted on that branch.
Once \(E(\Delta t)\) has been extracted from the boundary entropy data, its
inverse gives \(\Delta t(E)\), and the inverse problem reduces to solving the integral
equation
\begin{equation}
\Delta t(E)
=
2E\int_{0}^{+\infty}
\frac{dr}
{f(r)\sqrt{r^{2d-4}f(r)+E^2}} .
\label{eq:master_reconstruction_general}
\end{equation}
We use the time-width data rather than the entropy integral itself because \(\Delta t(E)\) is UV finite,
whereas the spacelike area contains the usual UV divergence.

\subsection{Abel reformulation in real coordinates}
\label{sec:abel_reformulation_real}
We now recast Eq.~\eqref{eq:master_reconstruction_general} as a
square-root-kernel integral transform whose discontinuity yields a standard
Abel integral equation on a single real radial branch. Define
\begin{equation}
W(r)\equiv r^{2d-4}f(r).
\label{eq:W-def-real}
\end{equation}
In terms of this variable, Eq.~\eqref{eq:master_reconstruction_general} becomes
\begin{equation}
\frac{\Delta t(E)}{2E}
=\int_{0}^{+\infty}
\frac{r^{2d-4}}{W(r)\sqrt{W(r)+E^{2}}}\,dr.
\end{equation}

The function \(W(r)\) is the transformed variable on the chosen real branch.
We impose \(W'(r)>0\) for \(r>0\) on this branch.  This
monotonicity condition makes \(W(r)\) strictly increasing and gives a
differentiable inverse \(r(W)\).  We also assume that the endpoint
\begin{equation}
    W_0\equiv \lim_{r\to0^+}W(r)
    \label{eq:W0_def_real}
\end{equation}
is finite.  The asymptotically AdS behavior
\(W(r)\sim r^{2d-2}\) implies \(W(r)\to+\infty\) as \(r\to\infty\).  If the
monotonicity condition fails, the radial integral must be split into locally
invertible pieces, and the corresponding branch contributions to \(H(W)\)
must be summed.

The real radial integral may cross a horizon, where \(f(r)=0\) and hence
\(W=0\). Whenever this happens, the integral above and all subsequent
\(W\)-integrals are defined with one fixed pole prescription. In the BTZ
example below we use the Cauchy-principal-value prescription on the real
branch. More generally, one may choose a fixed contour indentation around
the horizon pole; the same choice must then be used in the defining
time-width integral, the analytic continuation, and the Abel inversion.
Changing variables from \(r\) to \(W\) on this monotonic
branch gives
\begin{equation}
\frac{\Delta t(E)}{2E}
=\int_{W_{0}}^{+\infty}\frac{H(W)}{\sqrt{W+E^{2}}}\,dW,
\label{eq:abel-real-pre}
\end{equation}
where we have defined the density function
\begin{equation}
H(W)\equiv \frac{r(W)^{2d-4}}{W}\frac{dr}{dW}.
\label{eq:H-def-real}
\end{equation}
This is a square-root-kernel integral transform. The factor \(1/W\) in
\eqref{eq:H-def-real} means that \(H(W)\) may have a simple pole at a
horizon.  This pole is not an additional assumption; it is the same pole
already present in the original \(1/f(r)\) integrand.

To invert the transform, introduce the input-plane variable \(x=-E^2\) and
define the input function from the boundary time-width data by
\begin{equation}
\mathcal{I}(x)\equiv \frac{\Delta t(E)}{2E},
\qquad x=-E^2.
\label{eq:I-on-shell-real}
\end{equation}
For physical data with \(E^2>-W_0\), this determines \(\mathcal{I}(x)\) on the
real interval \(x<W_0\), namely on the real axis to the left of the branch cut
\([W_0,\infty)\).  The inverse problem is to find \(H(W)\) such that
\begin{equation}
\mathcal{I}(x)=\int_{W_{0}}^{+\infty}\frac{H(W)}{\sqrt{W-x}}\,dW,
\label{eq:I-def-real}
\end{equation}
where the square root is fixed by requiring \(\sqrt{W-x}>0\) for real
\(x\) and \(W\) satisfying \(x<W_0<W\).  After the shifts \(y=W-W_0\) and
\(z=W_0-x\), Eq.~\eqref{eq:I-def-real} is a generalized Stieltjes transform
of order \(\rho=1/2\)~\cite{ByrneLove1974,Schwarz2005}.  To reduce this
transform to the standard Abel form, we assume that the input function \(\mathcal I(x)\) admits an
analytic continuation to a neighborhood of the cut.
We further assume that \(W_0\) can be identified by matching the first
nonanalytic point of the input \(\mathcal I(x)\) to the
endpoint branch point of the Abel kernel.  This identification requires that no
earlier internal singularity, branch change, or endpoint cancellation masks the
endpoint; if the input data do not resolve it reliably, \(W_0\) must instead be
supplied independently or treated as a fitted parameter.

Let \(s\) lie on an interval of the cut that does not contain an internal
singularity of \(H\).  To evaluate the two boundary values explicitly, first
split the integral at \(W=s\):
\begin{equation}
\begin{aligned}
\mathcal I(s\pm i0)
={}&\int_{W_0}^{s}
\frac{H(W)}{\sqrt{W-s\mp i0}}\,dW
+\int_s^{+\infty}
\frac{H(W)}{\sqrt{W-s\mp i0}}\,dW.
\end{aligned}
\label{eq:I-boundary-split-real}
\end{equation}
For \(W>s\), the argument \(W-s\) is positive, so the second integral has the
same boundary value on both sides of the cut and cancels from the
discontinuity.  For \(W<s\), the branch fixed below Eq.~\eqref{eq:I-def-real}
gives
\begin{equation}
\begin{aligned}
W-(s+i0)&=-(s-W)-i0,
&\sqrt{W-s-i0}&=-i\sqrt{s-W},\\
W-(s-i0)&=-(s-W)+i0,
&\sqrt{W-s+i0}&=+i\sqrt{s-W}.
\end{aligned}
\label{eq:sqrt-boundary-values-real}
\end{equation}
Consequently,
\begin{equation}
\begin{aligned}
\frac{1}{\sqrt{W-(s+i0)}}&=\frac{i}{\sqrt{s-W}},
&
\frac{1}{\sqrt{W-(s-i0)}}&=-\frac{i}{\sqrt{s-W}}.
\end{aligned}
\label{eq:inverse-sqrt-boundary-values-real}
\end{equation}
Subtracting the lower boundary value from the upper one therefore yields
\begin{equation}
\mathrm{Disc}\,\mathcal{I}(s)
\equiv
\mathcal{I}(s+i0)-\mathcal{I}(s-i0)
=
2i\int_{W_{0}}^{s}\frac{H(W)}{\sqrt{s-W}}\,dW.
\label{eq:disc-real}
\end{equation}

Equation~\eqref{eq:disc-real} is the standard Abel integral equation, with
\(\mathrm{Disc}\,\mathcal I(s)/(2i)\) as the input, \(H(W)\) as the unknown
density function,\footnote{Throughout this paper, we refer to \(H(W)\) as the
Abel density.} and \((s-W)^{-1/2}\) as the integral
kernel.
The Abel density \(H(W)\) is then recovered by applying the standard Abel
inversion formula to
Eq.~\eqref{eq:disc-real}~\cite{PolyaninManzhirov2008}:
\begin{equation}
H(W)=\frac{1}{2\pi i}\frac{d}{dW}
\int_{W_{0}}^{W}
\frac{\mathrm{Disc}\,\mathcal{I}(s)}{\sqrt{W-s}}\,ds.
\label{eq:H-inverse-real}
\end{equation}
This inversion is local on the chosen \(W\)-branch. In particular, when
the analytic continuation of \(\mathcal I\) indicates isolated internal
singular points on the cut, the reconstruction is first performed on the open
intervals between them.  If a fixed prescription across such a point is
supplied, the reconstructed Abel density is then continued with the same
prescription.  

Once \(H(W)\) is known, the relation between \(W\) and the radial coordinate
follows from Eq.~\eqref{eq:H-def-real}.  Indeed, we have
\begin{equation}
\frac{d}{dW}\!\left(r^{2d-3}\right)
=(2d-3)\,H(W)W,
\label{eq:r-eq-real}
\end{equation}
which motivates the definition
\begin{equation}
G_{d}(W)\equiv (2d-3)\int_{W_{0}}^{W}\xi H(\xi)\,d\xi + C.
\label{eq:Gd-real}
\end{equation}
The Abel inversion determines \(H(W)\), but reconstructing the radial profile
requires the additional integration in Eq.~\eqref{eq:Gd-real}.  Identifying \(W_0\) from the input nonanalyticity is separate
from fixing this radial-origin ambiguity.  If \(W_0\) corresponds to the
black-hole singularity and the radial origin is chosen at that singularity,
the condition \(r(W_0)=0\) fixes \(C=0\).
Using the radial profile
\begin{equation}
r(W)=\bigl[G_{d}(W)\bigr]^{1/(2d-3)},
\label{eq:rW-real}
\end{equation}
the blackening factor is finally reconstructed as
\begin{equation}
f(r)=\frac{W(r)}{r^{2d-4}}.
\label{eq:f-final-real}
\end{equation}

Equations~\eqref{eq:abel-real-pre}--\eqref{eq:f-final-real} define the inverse
framework on any real branch satisfying the monotonicity and
endpoint conditions above.
For a closed-form demonstration, we specialize to the BTZ case with \(d=2\), whose
analytic TEE permits explicit inversion and an endpoint check.

\paragraph{Analytic Abel inversion in the BTZ case}

For $d=2$, one simply has
\begin{equation}
W(r)=r^{2d-4}f(r)=f(r),
\end{equation}
For the BTZ benchmark, \(f(r)=r^2-1\) gives
\(W'(r)=2r>0\) for \(r>0\), so the monotonicity condition required by the
Abel reduction is satisfied.
Equation~\eqref{eq:master_reconstruction_general} reduces to
\begin{equation}
\Delta t(E)=2E\int_{0}^{+\infty}\frac{dr}{f(r)\sqrt{f(r)+E^{2}}}.
\label{eq:btz-master-real}
\end{equation}
From the boundary TEE, one obtains
\begin{equation}
E(\Delta t)=-\coth\!\left(\frac{\Delta t}{2}\right),
\end{equation}
and hence
\begin{equation}
\Delta t(E)=2\,\mathrm{arccoth}(-E),
\qquad E\in(-\infty,-1).
\label{eq:btz-TE-real}
\end{equation}

The endpoint \(W_0\) is not inserted from the known BTZ metric.  Instead, it is
read from the analytic structure of the boundary input.
Introducing the Abel input variable \(x=-E^2\), the physical data therefore
lie on the real interval
\begin{equation}
E<-1
\quad\Longrightarrow\quad
E^2>1
\quad\Longrightarrow\quad
x=-E^2<-1.
\label{eq:btz-physical-x-range}
\end{equation}
At late time this interval terminates at
\begin{equation}
\Delta t\to\infty
\quad\Longrightarrow\quad
E\to-1^{-}
\quad\Longrightarrow\quad
E^2\to1^{+}
\quad\Longrightarrow\quad
x\to-1^{-}.
\label{eq:btz-late-time-map}
\end{equation}

We now pass to the Abel transform. 
For \(d=2\), the change of variables and the Abel density are
\begin{equation}
W=f(r),\qquad
W_0=\lim_{r\to0^+}W(r),\qquad
H(W)=\frac{1}{W}\frac{dr}{dW}.
\label{eq:btz-W-H-definitions}
\end{equation}
Substituting Eq.~\eqref{eq:btz-W-H-definitions} into
Eq.~\eqref{eq:btz-master-real} yields
\begin{equation}
\mathcal I(x)
\equiv\frac{\Delta t(E)}{2E}
=\int_{W_0}^{+\infty}
\frac{H(W)}{\sqrt{W-x}}\,dW,
\qquad x=-E^2.
\label{eq:btz-abel-input-general}
\end{equation}
Equation~\eqref{eq:btz-abel-input-general} is the BTZ specialization of the
general transform \eqref{eq:I-def-real}.  We now locate its endpoint
nonanalyticity directly from the closed-form boundary data.  On the physical
branch, \(E=-\sqrt{-x}\), with
\(\sqrt{-x}>0\) for real \(x<-1\).  Substituting
Eq.~\eqref{eq:btz-TE-real} gives
\begin{equation}
\mathcal I(x)
=\left.
\frac{2\operatorname{arccoth}(-E)}{2E}
\right|_{E=-\sqrt{-x}}
=-\frac{\operatorname{arccoth}\sqrt{-x}}{\sqrt{-x}}.
\label{eq:btz-input-closed-form}
\end{equation}
Here
\(\operatorname{arccoth}y=\frac12\log[(y+1)/(y-1)]\) for \(y>1\).
One can see that \(x=-1\) is a genuine logarithmic nonanalytic point, rather than merely
the end of the real parameterization.  Under the endpoint-identification
assumption stated after Eq.~\eqref{eq:I-def-real}, it fixes
\begin{equation}
W_0=-1.
\label{eq:btz-W0}
\end{equation}
{This is the singularity endpoint of the radial branch.}

It remains to compute the discontinuity needed for the Abel inversion.  Let
\(-1<s<0\) and set \(a=\sqrt{-s}\in(0,1)\).  Then
the square-root boundary values are
\begin{equation}
\sqrt{-(s+i0)}=a-i0,
\qquad
\sqrt{-(s-i0)}=a+i0.
\label{eq:btz-sqrt-boundary-values}
\end{equation}
Define \(R(y)=(y+1)/(y-1)\).  Since \(0<a<1\), one has
\(R(a)<0\) and \(R'(a)=-2/(a-1)^2<0\).  The negative derivative reverses the
side from which the square-root argument approaches, so
\begin{equation}
R_{\pm}
\equiv
\frac{\sqrt{-(s\pm i0)}+1}{\sqrt{-(s\pm i0)}-1}
=R(a)\pm i0.
\label{eq:btz-R-boundary-values}
\end{equation}
On the principal logarithmic branch, the corresponding boundary values are
\begin{equation}
\log R_{\pm}=\log|R(a)|\pm i\pi.
\label{eq:btz-log-boundary-values}
\end{equation}
Equation~\eqref{eq:btz-input-closed-form} then gives
\begin{equation}
\mathcal I(s\pm i0)
=-\frac{1}{2a}\bigl[\log|R(a)|\pm i\pi\bigr].
\label{eq:btz-I-boundary-values}
\end{equation}
Subtracting the lower boundary value from the upper one finally yields
\begin{equation}
\mathrm{Disc}\,\mathcal I(s)
\equiv\mathcal I(s+i0)-\mathcal I(s-i0)
=-\frac{i\pi}{a}
=-\frac{i\pi}{\sqrt{-s}}.
\label{eq:btz-disc}
\end{equation}

To proceed, we substitute this discontinuity into Eq.~\eqref{eq:disc-real}, which gives
\begin{equation}
\int_{-1}^{s}\frac{H(W)}{\sqrt{s-W}}\,dW
=
-\frac{\pi}{2\sqrt{-s}}.
\label{eq:btz-abel}
\end{equation}
Applying the inversion formula \eqref{eq:H-inverse-real}, we arrive at
\begin{equation}
H(W)=\frac{1}{2W\sqrt{1+W}}.
\label{eq:btz-H}
\end{equation}
This expression is obtained first for \(-1<W<0\).  On the chosen square-root
sheet it is analytic away from the branch point \(W=-1\) and has a simple pole
at the horizon \(W=0\). {The same expression gives the continuation to the exterior side \(W>0\).}

The remaining reconstruction is immediate.
Substituting Eq.~\eqref{eq:btz-H} into Eq.~\eqref{eq:Gd-real} and Eq.~\eqref{eq:rW-real} with $d=2$, we find
\begin{equation}
r(W)=\int_{-1}^{W}\frac{d\xi}{2\sqrt{1+\xi}} + C
=\sqrt{1+W}+C.
\end{equation}
{Applying the singularity condition \(r(W=-1)=0\) discussed above fixes the radial-origin ambiguity, yielding \(C=0\).}  Thus
\begin{equation}
r=\sqrt{1+W},
\qquad
W=r^{2}-1.
\end{equation}
Finally, because \(W=f\) for \(d=2\), the reconstructed blackening factor is
\begin{equation}
f(r)=r^{2}-1.
\label{eq:btz-final-f-real}
\end{equation}
Thus the selected-branch Abel inversion reproduces the BTZ metric function exactly.

\section{Complex-coordinate reconstruction}
\label{sec:3}

\subsection{Complex-coordinate prescription for HTEE}

We now formulate HTEE following the complex-extremal-surface framework of Refs.~\cite{Heller_2025,heller2025temporalentanglementholographicentanglement}.
We work in
\((d+1)\)-dimensional asymptotically AdS backgrounds
with \(d\ge2\) and allow an additional
spatial warp factor, so that the BTZ, Schwarzschild, RN, and
Gubser--Rocha examples can be discussed as special cases of the same notation.

The starting point is that the bulk dual of HTEE is not, in general, a real
extremal surface in the original Lorentzian geometry.  It is a complex
codimension-two extremal surface anchored on a timelike boundary subregion.
Both the embedding functions and the bulk contour can therefore be complex, and
Eq.~\eqref{eq:RT_formula} naturally gives a complex-valued entropy.

We use the inverse radial coordinate \(z=1/r\), following the conventions of
the complex-coordinate literature, and consider the static asymptotically AdS
metric
\begin{equation}
ds^2=\frac{1}{z^2}\left(
-f(z)\,dt^2+\frac{dz^2}{f(z)}
+h(z)\left(d\vec{x}^{\,2}+dy^2\right)
\right).
\label{eq:3.1_metric_generalized}
\end{equation}
Here \(z=0\) is the asymptotic boundary, \(f(z)\) is the blackening factor, and
\(h(z)\) describes a possible deformation of the spatial part of the metric.

We consider a strip-shaped timelike boundary subregion extended in the
\(\vec{x}\) directions, with \(y=0\).  By translational invariance along the
strip, the extremal surface has only one nontrivial embedding function.  We
parametrize it by \(z=z(t)\), so that the induced metric is
\begin{equation}
ds^2_{\mathrm{ind}}
=
\frac{1}{z^2}\left(
-f(z)+\frac{\dot z^{\,2}}{f(z)}
\right)dt^2
+\frac{h(z)}{z^2}d\vec{x}^{\,2}.
\label{eq:3.1_induced_metric}
\end{equation}
The corresponding area
functional can be written as
\begin{equation}
A
=
\int dt\,\mathcal{L}(z,\dot z),
\qquad
\mathcal{L}(z,\dot z)
=
\frac{h(z)^{(d-2)/2}}{z^{d-1}}
\sqrt{
-f(z)+\frac{\dot z^{\,2}}{f(z)}
}.
\label{eq:3.1_area_functional}
\end{equation}

Since the Lagrangian does not depend explicitly on \(t\), there is a conserved
quantity.  We denote it by \(E\):
\begin{equation}
E
=\mathcal{L}-
\dot z\,\frac{\partial \mathcal{L}}{\partial \dot z}
=
\frac{-h(z)^{(d-2)/2}f(z)}
{z^{d-1}\sqrt{-f(z)+\dot z^{\,2}/f(z)}}.
\label{eq:3.1_conserved_E}
\end{equation}
Solving this first integral for \(\dot z\), one obtains
\begin{equation}
\left(\frac{dt}{dz}\right)^2
=
\frac{E^2 z^{2d-2}}
{f(z)^2\left[h(z)^{d-2}f(z)+E^2 z^{2d-2}\right]}.
\label{eq:3.1_tprime}
\end{equation}

Equation~\eqref{eq:3.1_tprime} determines \(dt/dz\) only up to a local
square-root choice.  Following the path construction described in
Ref.~\cite{Heller_2025}, we define \(C\) as a contour representative
of a two-branch section of the complex extremal surface: it starts at the regulated boundary
\(z=\epsilon\) on the lower branch, reaches the selected square-root
branch point, passes to the other sheet, and returns to \(z=\epsilon\)
on the upper branch.  In the contour integrals below, the square roots
are fixed on the initial branch and analytically continued along \(C\).
The continuation to the other sheet implements the corresponding sign
change of \(dt/dz\).  Thus the choice of sign is encoded by the analytic
continuation along \(C\) and is not written separately in the contour
integrals below.  With these in mind, the time-width and area functionals
are written as
\begin{equation}
\Delta t(E)
=
\int_C dz\,
\frac{E z^{d-1}}
{f(z)\sqrt{h(z)^{d-2}f(z)+E^2 z^{2d-2}}},
\label{eq:3.1_Deltat}
\end{equation}
and
\begin{equation}
A(E)
=
\int_C dz\,
\frac{h(z)^{d-2}}
{z^{d-1}\sqrt{h(z)^{d-2}f(z)+E^2 z^{2d-2}}}.
\label{eq:3.1_area_onshell}
\end{equation}
Accordingly, the HTEE is
\begin{equation}
S_T(E)=\frac{A(E)}{4}.
\label{eq:3.1_ST_def}
\end{equation}

Along a fixed extremal-surface branch, we may choose a slightly enlarged representative
of \(C\) such that the branch point remains strictly inside the contour under
an infinitesimal variation of \(E\), without any pole or branch cut crossing
the contour. The contour can therefore be held fixed during differentiation,
so only the explicit \(E\)-dependence of the integrands contributes. Equivalently, the following derivatives are the explicit form
of the on-shell variation. Differentiating Eqs.~\eqref{eq:3.1_Deltat} and
\eqref{eq:3.1_area_onshell} with respect to \(E\), we obtain
\begin{align}
\frac{d\Delta t}{dE}
&=
\int_C dz\,
\frac{\partial}{\partial E}
\left[
\frac{E z^{d-1}}
{f(z)\sqrt{h(z)^{d-2}f(z)+E^2z^{2d-2}}}
\right]
\notag\\
&=
\int_C dz\,
\frac{h(z)^{d-2}z^{d-1}}
{\left[h(z)^{d-2}f(z)+E^2z^{2d-2}\right]^{3/2}},
\label{eq:complex_dDelta_dE}\\[2mm]
\frac{dS_T}{dE}
&=
\frac{1}{4}
\int_C dz\,
\frac{\partial}{\partial E}
\left[
\frac{h(z)^{d-2}}
{z^{d-1}\sqrt{h(z)^{d-2}f(z)+E^2z^{2d-2}}}
\right]
\notag\\
&=
-\frac{E}{4}
\int_C dz\,
\frac{h(z)^{d-2}z^{d-1}}
{\left[h(z)^{d-2}f(z)+E^2z^{2d-2}\right]^{3/2}} .
\label{eq:complex_dS_dE}
\end{align}
The same contour integral therefore cancels in the ratio, giving
\begin{equation}
    \frac{dS_T}{d\Delta t}
    =
    \frac{dS_T/dE}{d\Delta t/dE}
    =
    -\frac{E}{4}.
    \label{eq:complex_HJ}
\end{equation}
The UV subtraction used below is independent of \(E\), so the same relation
holds for \(S_{\rm reg}\).

In the complex-coordinate prescription, the time-width
and area integrals share the same square-root denominator.  For fixed \(E\),
this denominator may have several zeros in the complex \(z\)-plane, which are
generically branch points of the integrands.  We denote by \(z_t\) the zero
selected by \(C\), where the contour passes from the lower to the upper branch,
and refer to it as the turning point.  It satisfies
\begin{equation}
E^2=-\frac{h(z_t)^{d-2}f(z_t)}{z_t^{2d-2}}.
\label{eq:3.1_turning_point}
\end{equation}
The remaining zeros constrain the cut structure and the
admissible homotopy classes of contours.

The contour \(C\) is not unique.  For fixed \(E\) and a selected
turning point \(z_t\), any two-branch representative in the same homotopy class
gives the same contour integrals, provided the deformation avoids poles and
other branch points and remains on the chosen square-root sheet.  We use
standard complex-analytic conventions for contour deformation, branch cuts,
and analytic continuation~\cite{AblowitzFokas2003,Forster1981}.  The complex
extremal surface is therefore represented by an equivalence class of such
contours, rather than by a single distinguished path in the complex
\(z\)-plane.

The late-time behavior is governed by a critical extremal surface, defined by stationarity of its constant-\(z\) area
density~\cite{HartmanMaldacena2013,Heller_2025}.  Equivalently, the associated
branch point becomes degenerate, which occurs when
\begin{equation}
\left.
\partial_z\!\left(\frac{h(z)^{d-2}f(z)}{z^{2d-2}}\right)
\right|_{z=z_c}=0.
\label{eq:3.1_critical_point}
\end{equation}
At such a point,
\begin{equation}
E_c^2=-\frac{h(z_c)^{d-2}f(z_c)}{z_c^{2d-2}},
\label{eq:3.1_Ec}
\end{equation}
and the corresponding value of the on-shell Lagrangian controls the late-time
linear growth of the entropy.  This critical behavior will be used in the RN
and Gubser--Rocha applications below.

To summarize, the complex-coordinate formalism in the generalized metric
\eqref{eq:3.1_metric_generalized} provides a common forward description for
backgrounds with either a trivial or a nontrivial spatial warp factor. The next
subsection uses this framework to compute \(\Delta t(E)\) and \(S_T(E)\).

\subsection{Forward computation method}
\label{subsec:forward_method}

We use the SAdS\(_4\) geometry as a representative example to illustrate the
forward computation in detail. The same numerical procedure is later applied
to the RN and Gubser--Rocha geometries, with only the corresponding blackening
and warp factors modified.

For the SAdS\(_4\) test geometry we set
\begin{equation}
    f(z)=1-z^3 .
    \label{eq:f_ads4_sch_forward}
\end{equation}
The turning-point condition gives
\begin{equation}
    E^2=\frac{z_t^3-1}{z_t^4}.
    \label{eq:E_zt_forward}
\end{equation}
Solving for $E$ requires a branch choice. For the upper vacuum-connected branch traced below, we implement a compatible choice at the vacuum endpoint by requiring
\(\sqrt{1-z_t^3}\to+1\) as \(z_t\to0\) and choosing the sign of \(E\) so that,
in the vacuum limit, positive \(\Delta t\) corresponds to
\(\operatorname{Im}z_t>0\).  Thus,
\begin{equation}
    E=-\,\frac{i\sqrt{1-z_t^3}}{z_t^2}.
    \label{eq:E_zt_branch_forward}
\end{equation}
The square root, and hence \(E\), is then tracked continuously along the
branch.

It is useful to introduce
\begin{equation}
    g(z;z_t)
    \equiv f(z)+E^2z^4
    =
    1-z^3+\frac{z^4(z_t^3-1)}{z_t^4}.
    \label{eq:g_def_forward}
\end{equation}
The zero of \(g(z;z_t)\) at \(z=z_t\) is the square-root branch point of the
extremal surface. In terms of
\(z_t\), the boundary time separation and regularized entropy are written as
\begin{equation}
    \Delta t(z_t)
    =
    2\int_{\bar C(\epsilon,z_t)}
    \frac{i z^2\sqrt{1-z_t^3}}
    {(z^3-1)z_t^2\sqrt{g(z;z_t)}}\,dz ,
    \label{eq:t_numeric_forward}
\end{equation}
and
\begin{equation}
    S_{\rm reg}(z_t)
    =
    2\times\frac{1}{4}
    \int_{\bar C(\epsilon,z_t)}
    \frac{dz}{z^2\sqrt{g(z;z_t)}}
    -\frac{1}{2\epsilon}.
    \label{eq:Sreg_numeric_forward}
\end{equation}
Here \(\bar C(\epsilon,z_t)\) denotes a one-way representative of the
two-branch contour class defined above, running from the regulated boundary
\(z=\epsilon\) to the selected turning point \(z_t\).  We use the same
regulated lower endpoint for both observables.\footnote{Near the asymptotic
boundary, \(g(z;z_t)=1+O(z^3)\), so the time-width integrand is \(O(z^2)\).
Replacing its lower endpoint \(\epsilon\) by \(0\) therefore changes
\(\Delta t\) only by \(O(\epsilon^3)\). By contrast, the entropy integrand behaves as \(z^{-2}\), so
Eq.~\eqref{eq:Sreg_numeric_forward} must retain the cutoff. Its standard UV
divergence is removed by the explicit counterterm \(1/(2\epsilon)\).
The regulated formulas above retain \(z=\epsilon\) as a common
lower endpoint for uniform contour notation.  In the numerical implementation, the
time-width integral is evaluated from \(z=0\).  For the entropy, we subtract
the leading \(z^{-2}\) term at the integrand level, integrate the regular
remainder from \(z=0\), and retain the associated finite endpoint term
\(-1/z_t\).  This is algebraically equivalent to the \(\epsilon\to0\) limit of
Eq.~\eqref{eq:Sreg_numeric_forward} and avoids cancellation between
\(O(\epsilon^{-1})\) contributions.}
The full two-branch section continues through \(z_t\) onto the second sheet and
returns to the same boundary endpoint on the other branch.  Across the branch
point, the square root changes sign, while the orientation of the second branch
is opposite to that of the first one.  These two minus signs cancel, so the two
branches give equal contributions.  This is the origin of the explicit branch
factor of \(2\) in Eqs.~\eqref{eq:t_numeric_forward} and
\eqref{eq:Sreg_numeric_forward}.

At the turning-point endpoint, a nondegenerate root obeys
\(g(z;z_t)=g'(z_t;z_t)(z-z_t)+\cdots\), so the integrand behaves as
\((z-z_t)^{-1/2}\).  This square-root singularity is locally
integrable, but it is numerically stiff.  For the numerical quadrature of both
observables, we introduce the normalized arclength parameter
\(\lambda\in[0,1]\) along \(\bar C(\epsilon,z_t)\) and regard the contour as a
complex-valued map
\begin{equation}
    z=z(\lambda), \qquad \lambda\in[0,1],
\end{equation}
from the boundary point \(z(0)=\epsilon\) to the branch point
\(z(1)=z_t\).  For the final segment approaching
the endpoint \(z=z_t\), we stop the numerical integration at
\(\lambda=1-\epsilon_{\rm tp}\), rather than integrating directly to
\(\lambda=1\), and treat the remaining short endpoint contribution
\(\lambda\in[1-\epsilon_{\rm tp},1]\) separately as a tail correction. In this
tail segment the last contour direction is kept fixed and the integrand is
expanded locally in powers of \((z_t-z)^{1/2}\). The singular leading term is
then integrated analytically over the short final interval, while the regular
remainder is evaluated numerically. This implements the same contour but avoids
the loss of accuracy that would come from direct quadrature at the square-root
endpoint.

\paragraph{Contour optimization.}
For each value of \(z_t\), the contour \(\bar C(\epsilon,z_t)\) is discretized into a finite
set of nodes. Inspired by elastic-band methods in robotic path planning
\cite{QuinlanKhatib1993ElasticBands}, we determine the contour by an adaptive
elastic-band algorithm:
neighboring nodes are connected by a smoothing force, while the singular points
of the integrand act as repulsive obstacles. In the SAdS\(_4\)
case these obstacles include the zeros of \(f(z)\) and the additional zeros of
\(g(z;z_t)\), except for the endpoint root at \(z=z_t\).  Importantly, the contour
at each tracing step is initialized from the optimized contour at the previous
step and is then deformed continuously as \(z_t\) is advanced.  This provides a
practical way to transport the contour continuously between
successive values of \(z_t\) and helps prevent the numerical integration path
from jumping across poles or branch points.  After each relaxation step, the
contour is resampled with
a position-dependent node density, so that regions close to poles or branch
points are resolved more finely.  This procedure does not introduce
an additional physical assumption; rather, it selects a stable numerical contour
within this continuously tracked family while remaining on the chosen square-root
sheet.
The complex-extremal-surface framework of
Refs.~\cite{Heller_2025,heller2025temporalentanglementholographicentanglement}
identifies the relevant complex branches, but it does not provide a concrete
contour optimization algorithm; the elastic-band procedure above is the
explicit numerical prescription used in this work.

\begin{figure}[ht]
    \centering
    \includegraphics[width=0.62\textwidth]{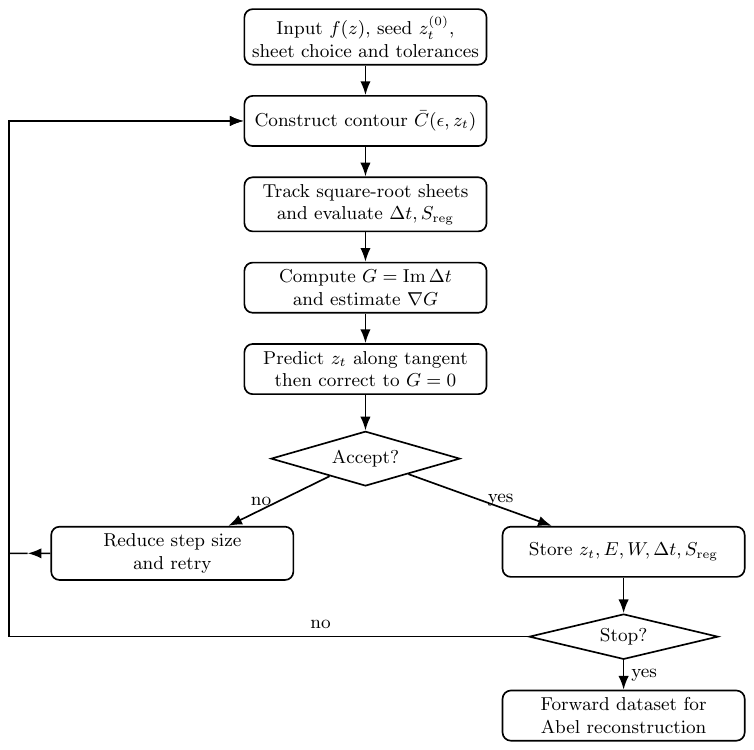}
    \caption{Flowchart of the forward computation.  Each update of the turning
    point \(z_t\) requires a new contour optimization, because the roots of
    \(g(z;z_t)\) and hence the branch structure of the integrand change along
    the traced branch.}
    \label{fig:forward_flowchart}
\end{figure}

\paragraph{Sheet and branch tracking.}
The square roots in \eqref{eq:t_numeric_forward} and
\eqref{eq:Sreg_numeric_forward} must be followed on a continuous sheet.  At each
node, the sign of the square root is fixed by continuity with the previous node,
\begin{equation}
    \sqrt{g(z_j;z_t)}
    \;\longrightarrow\;
    \sigma_j\sqrt{g(z_j;z_t)},\qquad
    \sigma_j=\operatorname*{arg\,min}_{\sigma=\pm1}
    \left|
    \sigma\sqrt{g(z_j;z_t)}-\sqrt{g(z_{j-1};z_t)}
    \right| .
    \label{eq:sqrt_tracking_forward}
\end{equation}
The same prescription is applied to \(\sqrt{1-z_t^3}\) between successive
values of \(z_t\) during branch continuation.  This prevents artificial jumps
between sheets of the Riemann surface.

With this continuous-sheet prescription, we trace the upper
vacuum-connected branch identified in Ref.~\cite{Heller_2025}.  For the purely
timelike SAdS\(_4\) strip---the
\(\theta=\pi/2\) case of
Ref.~\cite{heller2025temporalentanglementholographicentanglement}---this is
precisely the branch that determines the TEE under the prescription of that
reference: analytic continuation followed by minimization of the real part of
the area.
Its turning point approaches the asymptotic boundary in the
\(\Delta t\to0\) limit and therefore connects continuously to the vacuum
timelike-strip result.  Other complex branches can satisfy the same boundary
anchoring conditions, but they would define different candidate HTEE data and
hence different inverse reconstruction problems.

Real-boundary-time points on the selected branch satisfy
\begin{equation}
    \mathrm{Im}\,\Delta t(z_t)=0 .
    \label{eq:reality_condition_forward}
\end{equation}
Let
\begin{equation}
    G(z_t)\equiv \mathrm{Im}\,\Delta t(z_t),\qquad z_t=x+iy .
\end{equation}
The real-time locus on this branch is traced by solving \(G=0\). At a given point, the gradient is
estimated by finite differences,
\begin{equation}
    \partial_xG\simeq
    \frac{G(z_t+\delta)-G(z_t)}{\delta},\qquad
    \partial_yG\simeq
    \frac{G(z_t+i\delta)-G(z_t)}{\delta}.
    \label{eq:finite_diff_gradient_forward}
\end{equation}
Here \(\delta>0\) is a small real finite-difference step.
The tangent direction is chosen perpendicular to this gradient,
\begin{equation}
    \tau \propto i\nabla G ,
    \label{eq:tangent_forward}
\end{equation}
with the sign fixed by requiring \(\mathrm{Re}\,\Delta t\) to increase along the
branch. A trial step is then projected back to \(G=0\) by
\begin{equation}
    z_t^{\rm new}
    =
    z_t^{\rm trial}
    -
    \frac{G(z_t^{\rm trial})}
    {|\nabla G(z_t^{\rm trial})|^2}
    \nabla G(z_t^{\rm trial}) .
    \label{eq:newton_projection_forward}
\end{equation}
Steps for which \(|\mathrm{Im}\,\Delta t|\) becomes too large, for which
\(\mathrm{Re}\,\Delta t\) jumps discontinuously, or for which the contour
optimization fails are rejected and retried with a smaller step size.  In the
SAdS\(_4\) example shown below we follow the upper branch,
\(\mathrm{Im}\,z_t\ge0\); if a trial point crosses to the lower half-plane, the
solution is reflected by complex conjugation.

The resulting forward data are
\begin{equation}
    \left\{
    z_t(\lambda),\,
    E(\lambda),\,
    \Delta t(\lambda),\,
    S_{\rm reg}(\lambda)
    \right\},
    \label{eq:forward_dataset}
\end{equation}
where \(\lambda\) denotes the arclength-type tracing parameter along the
selected vacuum-connected branch, normalized to the sampled path. These
data will serve as the input for reconstruction by Abel inversion.

Figure~\ref{fig:forward_flowchart} summarizes the numerical pipeline.  The forward
problem is a coupled contour-tracing problem: after every update of \(z_t\), the
contour must be re-optimized before the time and entropy integrals can be
evaluated reliably.
Appendix~\ref{app:numerical_parameters} summarizes the principal numerical
parameters used in the forward and reconstruction algorithms and gives the
explicit elastic-band relaxation prescription used to optimize the tracked
complex contours.

\begin{figure}[!htbp]
    \centering
    \includegraphics[width=0.85\textwidth]{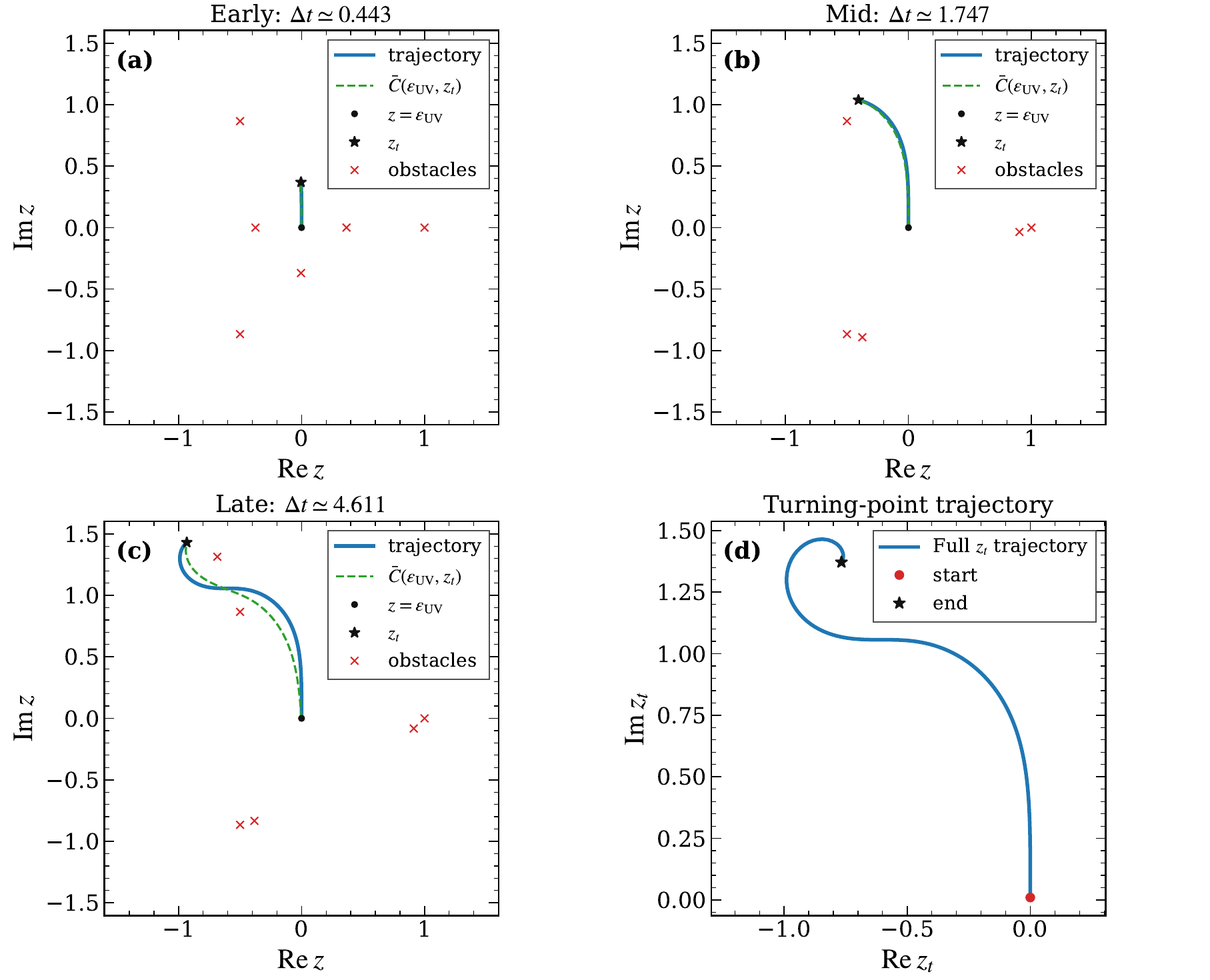}
    \caption{Contour tracing in the SAdS\(_4\) background.
    Panels (a)--(c) show three representative stages of the tracing procedure,
    together with the corresponding integration contours \(\bar C\) in the
    complex \(z\)-plane.  In these panels, the solid blue curve denotes the
    traced turning-point trajectory up to the current value of \(z_t\), the green
    dashed curve denotes the one-way integration contour \(\bar C(\epsilon,z_t)\), the
    black dot marks the boundary point, the black star marks the
    selected turning point \(z_t\), and the red crosses mark the singular points
    treated as repulsive obstacles in the contour relaxation.  Panel (d) shows
    the full trajectory of the selected turning point \(z_t\).  The snapshots illustrate that the contour deforms
    continuously as the turning point moves along the selected vacuum-connected
branch while satisfying \(\mathrm{Im}\,\Delta t=0\).}
    \label{fig:tracing_snapshots_and_trajectory}
\end{figure}

The contour snapshots in Fig.~\ref{fig:tracing_snapshots_and_trajectory} make
explicit how the selected branch is represented by a continuous family of
complex contours parametrized by the turning point \(z_t\).  As \(z_t\) moves
along the real-time locus \(\mathrm{Im}\,\Delta t=0\), the optimized contour is
continuously deformed from one tracing step to the next, without
switching to a different route around poles or branch points or leaving the
chosen square-root sheet.

\begin{figure}[!htbp]
    \centering
    \includegraphics[width=0.463\textwidth]{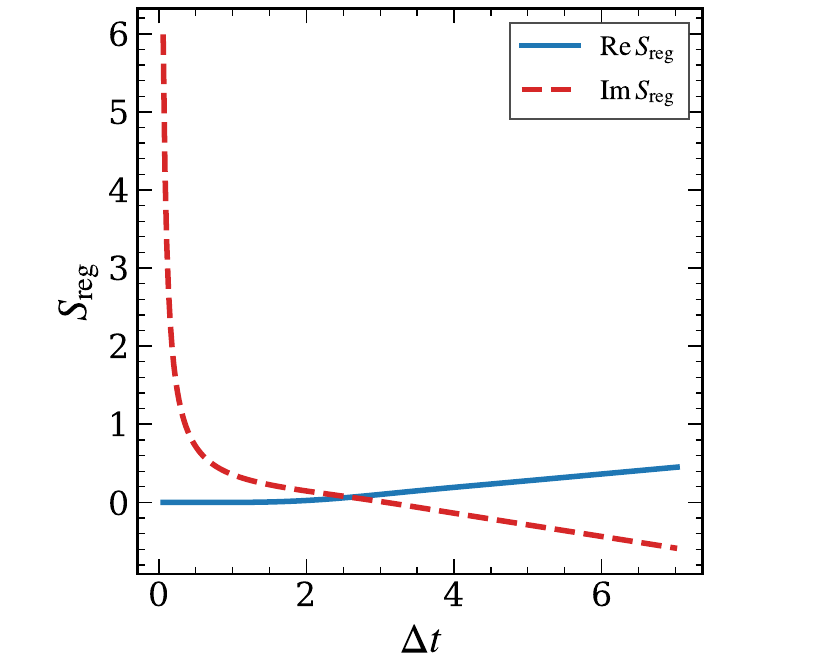}
    \caption{Regularized TEE, in the SAdS\(_4\) background.  The real and imaginary parts of
    \(S_{\rm reg}\) are shown as functions of the real boundary time
    \(\Delta t\) along the traced upper vacuum-connected branch.}
    \label{fig:ads4_forward_entropy}
\end{figure}

The corresponding entropy curve is shown in Fig.~\ref{fig:ads4_forward_entropy}.
The structure of the traced branch and the resulting complex entropy are in
agreement with the complex-extremal-surface picture of previous
works~\cite{Heller_2025,heller2025temporalentanglementholographicentanglement}.  The
forward data obtained in this way provide the input for the
inverse reconstruction considered below.

\subsection{Reconstruction by Abel inversion}
\label{subsec:abel_complex_coordinate}

We now reduce the time-width contour integral to an Abel integral equation.  This
subsection treats the one-function case \(h(z)=1\), in a
\((d+1)\)-dimensional metric
\begin{equation}
    ds^2=z^{-2}\left[-f(z)dt^2+\frac{dz^2}{f(z)}+d\vec x^{\,2}\right].
    \label{eq:metric_one_function_abel}
\end{equation}
The nontrivial-warp-factor case is not a direct corollary of this
one-function inversion; it is discussed separately in
Sec.~\ref{subsec:GR_application}.

\paragraph{Abel input and sheet prescription.}
In an inverse application the directly supplied boundary datum is the
regularized entropy curve \(S_{\rm reg}(\Delta t)\), with \(\Delta t\ge 0\).
A pre-labelled function \(\Delta t(E)\) is instead a derived representation of
these data. On a given extremal-surface branch,
the conserved parameter is fixed by the Hamilton--Jacobi relation
\begin{equation}
    E(\Delta t)=-{4}\frac{dS_{\rm reg}}{d\Delta t}.
    \label{eq:E_from_Sreg_branch_choice}
\end{equation}
Thus \(E\) is not subject to an additional branch choice after differentiating
\(S_{\rm reg}(\Delta t)\): the ordered curve \(E(\Delta t)\), and hence
\begin{equation}
    x(\Delta t)=-E(\Delta t)^2 ,
    \label{eq:x_from_Sreg_branch_choice}
\end{equation}
are data derived from the entropy curve on the selected branch.

The vacuum-connected branch is selected at the UV end.  We require that, as
\(\Delta t\to0\), the entropy curve approach the vacuum timelike-strip result and
\begin{equation}
    |E(\Delta t)|\to\infty .
    \label{eq:vc_condition_E}
\end{equation}
This condition selects the branch whose chosen turning point approaches the
asymptotic boundary in the UV limit.

The square-root sheet is fixed by a UV normalization.  At the regulated
boundary we choose
\begin{equation}
    \sqrt{f(z)+E^2z^{2d-2}}\longrightarrow +1 ,
    \qquad z\to0 ,
    \label{eq:uv_sqrt_sheet_choice}
\end{equation}
The sign in \eqref{eq:uv_sqrt_sheet_choice} is a normalization convention: it is the sheet
that reduces to the standard positive radial square root in the vacuum
near-boundary region.  Choosing the opposite sign would reverse the sign of the
square-root integral and must be accompanied by the corresponding reversal of the contour
orientation or of the \(E\)-branch; it would not define an independent
vacuum-connected datum.

The square-root factors are then continued continuously, in the
sense of analytic continuation on the selected sheet~\cite{Forster1981}, along the ordered
\(x(\Delta t)\) curve determined by the entropy data.  No
additional \(x\)-plane choice is needed when the inversion is implemented
directly on this curve.  Such a choice enters only if one evaluates
\(\mathcal I(x)\) away from the data path and has to pass a pole or branch
point.

\paragraph{Abel reduction and metric reconstruction.}
On a chosen one-way contour \(\bar C(\epsilon,z_t)\), half of the boundary time
width is
\begin{equation}
    \frac{\Delta t(E)}{2}
    =
    \int_{\bar C(\epsilon,z_t)}
    \frac{E\,z^{d-1}}
    {f(z)\sqrt{f(z)+E^2 z^{2d-2}}}\,dz.
    \label{eq:half_Deltat_E_general}
\end{equation}
To reduce this expression to an Abel integral equation, we introduce the
branch-dependent transformed variable
\begin{equation}
    W(z)=\frac{f(z)}{z^{2d-2}}.
    \label{eq:W_def_main}
\end{equation}
For an asymptotically AdS metric, \(f(z)\to1\) as \(z\to0\).  Hence
\(W(z)\sim z^{-(2d-2)}\), so \(W(\epsilon)\to\infty\) as
\(\epsilon\to0\). On the selected branch, the turning-point condition
\eqref{eq:3.1_turning_point} reduces, for \(h=1\), to
\(W(z_t)=-E^2\). 

We need a complex-coordinate analogue of the real-branch monotonicity condition given in Sec.~\ref{sec:2}. Following Ref.~\cite{EstradaKanwal2000}, let \(\Gamma\) be a
fixed directed simple smooth curve in the \(W\)-plane ending at the
regulated-boundary value \(W(\epsilon)\).  For \(x\in\Gamma\), let
\(\Gamma_{x,W(\epsilon)}\) denote the path from \(x\) to \(W(\epsilon)\) along
\(\Gamma\).  We assume that, for each \(E\), the selected one-way contour
\(\bar C(\epsilon,z_t)\) can be continuously deformed, with its endpoints fixed
and without crossing poles or other branch points, while remaining on the
tracked sheet, to a representative whose oppositely oriented \(W\)-image is
\(\Gamma_{x,W(\epsilon)}\).  A single-valued differentiable inverse
\(z(W)\) is required to exist along this path. In particular,
the representative must not pass through a critical point of \(W\), where
\(dW/dz=0\). If no such inverse exists on the whole path, the contour integral
must instead be treated piecewise, or the contour family must be truncated
before the critical point, as in the numerical applications below.

Under this assumption, the
time-width integral simplifies to
\begin{equation}
    \frac{\Delta t(E)}{2E}
    =
    \int_{-E^2}^{\infty}
    \frac{H(W)}{\sqrt{W+E^2}}\,dW,
    \label{eq:Deltat_over_2E_abel}
\end{equation}
where the Abel density \(H(W)\) is defined by
\begin{equation}
    H(W)
    \equiv
    -\frac{1}{z^{2d-2}W}\frac{dz}{dW}.
    \label{eq:H_def_first}
\end{equation}
If we further introduce \(x=-E^2\) and the input function
\begin{equation}
    \mathcal{I}(x)\equiv \frac{\Delta t(E)}{2E},
    \label{eq:abel_input_complex}
\end{equation}
Eq.~\eqref{eq:Deltat_over_2E_abel} becomes
\begin{equation}
    \mathcal{I}(x)
    =
    \int_{x}^{\infty}
    \frac{H(W)}{\sqrt{W-x}}\,dW.
    \label{eq:abel_standard_main}
\end{equation}
This is a complex-contour Abel integral equation involving the adjoint Abel operator~\cite{EstradaKanwal2000}.
The value \(x=-E^2\) is evaluated on
the selected extremal-surface branch; it is not
a new independent branch choice.  The input function \(\mathcal{I}(x)\) is
interpreted with the ordered branch data described above.  The inversion of
\eqref{eq:abel_standard_main} uses the endpoint coordinate \(W_t=x=-E^2\)
supplied by the boundary data; knowledge of the map \(z\mapsto W(z)\) is not
needed to perform the inversion itself.  That map is reconstructed only after
the Abel density has been obtained and a geometric anchor has been supplied.

The corresponding inversion formula can be read as~\cite{EstradaKanwal2000}
\begin{equation}
    H(W)
    =
    -\frac{1}{\pi}
    \frac{d}{dW}
    \int_{W}^{\infty}
    \frac{\mathcal{I}(x)}
    {\sqrt{x-W}}\,dx.
    \label{eq:abel_inverse_main}
\end{equation}
Here \(W\) and \(x\) lie on the fixed directed simple smooth
curve \(\Gamma\), and the integral is taken along the path
\(\Gamma_{W,W(\epsilon)}\) from \(W\) to \(W(\epsilon)\), with
\(W(\epsilon)\to\infty\) in the boundary limit.  The square root is understood on
the sheet fixed above.
Once \(H(W)\) has been determined from the time-width data, the radial profile
is obtained by one further integration:
\begin{equation}
    z^{-(2d-3)}(W)
    =
    (2d-3)\int \xi H(\xi)\,d\xi + C
    \equiv
    G_d(W).
    \label{eq:Gd_def_main}
\end{equation}

As in the real-coordinate reconstruction of Sec.~\ref{sec:2}, Abel inversion
determines the density \(H(W)\) but leaves the additive constant
\(C\) in \eqref{eq:Gd_def_main} unfixed.  In the complex-coordinate
prescription, the extremal surface is continued along a complex radial path and,
unlike the selected CWES branch used in Sec.~\ref{sec:2}, is not anchored at
the black-hole singularity.
Thus no built-in condition fixes \(C\).  One additional geometric input, such as
the horizon position or a residual near-boundary coefficient
introduced below, must be supplied; without it, the complex-coordinate
reconstruction by Abel inversion determines a family of blackening-factor profiles rather
than a unique one.

Finally, the original metric function is reconstructed through
\begin{equation}
    f(z)=z^{2d-2}W(z).
    \label{eq:f_reconstruct_main}
\end{equation}
The spacetime dimension affects the final branch selection.  For \(d=2\),
\(z(W)=1/G_2(W)\), so the radial coordinate is
single-valued once \(C\) is
fixed.  For \(d=3\), \(z(W)=[G_3(W)]^{-1/3}\); the reconstructed path therefore
lives on a three-sheeted Riemann surface.  We select the root branch reached by
continuous continuation from the regulated boundary, where \(z\) is the positive
real AdS radial coordinate and \(z(W)\sim W^{-1/(2d-2)}\).

\paragraph{Analytic reconstruction for the BTZ black hole}
\label{subsubsec:btz_analytic_reconstruction}

For the BTZ black hole, the reconstruction
can be carried out analytically and checks the formalism described above.

Consider the physical input function
\begin{equation}
    \mathcal{I}(-E^2)
    =
    \frac{\Delta t(E)}{2E}
    =
    \frac{\operatorname{arccoth}(-E)}{E}.
    \label{eq:btz_I_input}
\end{equation}
The \(\operatorname{arccoth}\) sheet is chosen so that
\(E=-\coth(\Delta t/2)\) for \(\Delta t>0\).  Hence
\(E\sim-2/\Delta t\) as \(\Delta t\to0\), and the endpoint relation
\(W(z_t)=-E^2\) gives \(|W(z_t)|\to\infty\).  Since
\(W(z)\sim z^{-2}\) near the AdS boundary for \(d=2\),
the selected turning point approaches \(z_t\to0\), so
\eqref{eq:btz_I_input} is the vacuum-connected input function.

To perform the inversion, define
\begin{equation}
    \mathcal{J}(W)\equiv
    \int_{W}^{\infty}\frac{\mathcal{I}(x)}{\sqrt{x-W}}\,dx .
\end{equation}
Here the upper endpoint \(\infty\) is the image of the asymptotically AdS
boundary.
For real \(E<-1\), the physical turning point has
\(W(z_t)=-E^2<-1\).  It is convenient, however, to first evaluate the Abel
density in the region \(W>0\), where the integration contour does not cross the
horizon pole at \(W=0\), and then analytically continue the result to the
physical branch.
Thus the continuation used below is not a second choice of the physical
\(E(\Delta t)\) branch; it is an auxiliary analytic continuation away from the
physical data path \(x=-E^2<-1\), introduced to obtain a closed-form BTZ
evaluation.

The Abel integral for \(\mathcal J(W)\) therefore runs along the positive real \(x\)-axis and uses the upper boundary value of the continued input function there. The corresponding value of \(E\) must be obtained by continuing the physical vacuum-connected germ. On the physical data path,
\begin{equation}
    E=-\sqrt{-x},\qquad \sqrt{-x}>0\quad (x<-1),
    \label{eq:btz_E_physical_germ}
\end{equation}
where the sign is fixed by \(E<0\).

\begin{figure}[!htbp]
    \centering
    \includegraphics[width=\textwidth]{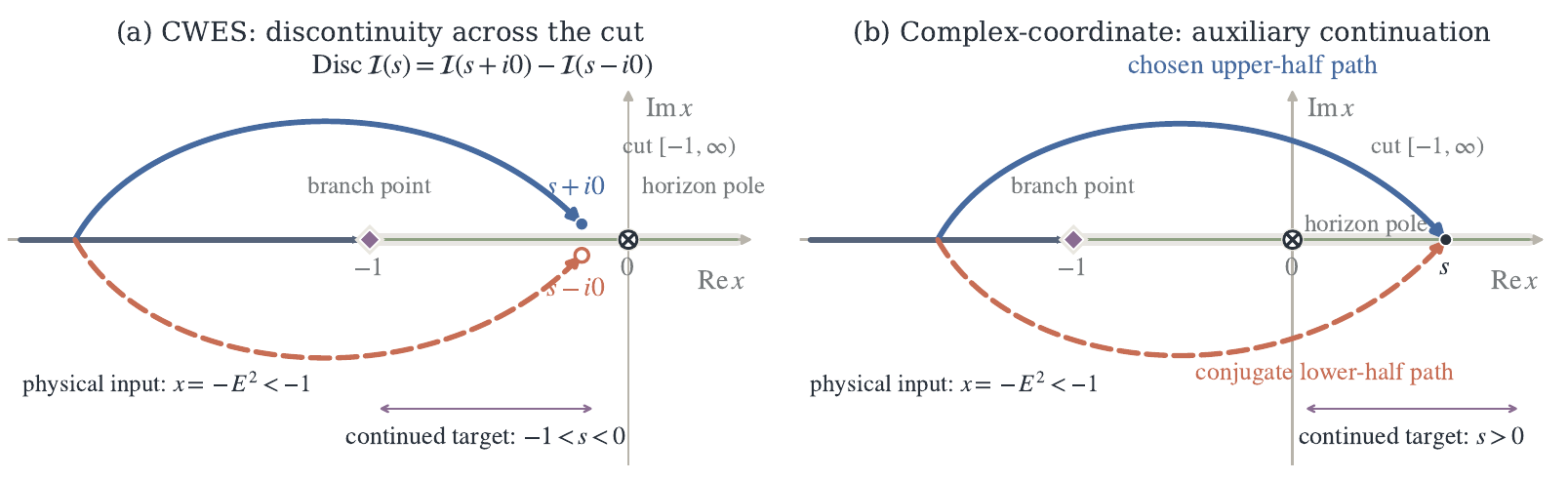}
    \caption{Analytic-continuation paths in the two BTZ reconstructions.
    (a) The Abel input for the CWES-based reconstruction is the discontinuity between the boundary values at
    \(s\pm i0\) across the cut, with \(-1<s<0\).
    (b) In the complex-coordinate reconstruction, the physical germ \(x=-E^2<-1\) is
    continued through the upper half \(x\)-plane to \(s>0\); the dashed lower
    path is its conjugate and gives the same \(\mathcal I(s)\)
    on the positive axis.}
    \label{fig:btz_continuation_paths}
\end{figure}

To specify the continuation explicitly, we analytically continue \(x\) from the negative real axis to a target point \(x=s\) (\(s>0\)) on the positive real axis along a path in the upper half-plane (\(\operatorname{Im}x>0\)). Figure~\ref{fig:btz_continuation_paths} contrasts this auxiliary continuation
with the cut discontinuity used in the CWES-based reconstruction of
Sec.~\ref{sec:abel_reformulation_real}.

Under this continuation, the continuous argument of \(-x\) decreases from \(0\) to \(-\pi\), so that \(-x \to s e^{-i\pi}\). Consequently, the positive real axis is approached from above with
\begin{equation}
    \sqrt{-x} \to \sqrt{s}\,e^{-i\pi/2} = -i\sqrt{s}.
    \label{eq:btz_upper_sqrt_endpoint}
\end{equation}
The conserved parameter therefore continues to
\begin{equation}
    E=i\sqrt{s},
    \label{eq:btz_E_continued}
\end{equation}
on the positive real axis. The \(\operatorname{arccoth}\) branch is continued along
the same path. On the resulting sheet,
\begin{equation}
    \operatorname{arccoth}(-i\sqrt{s})
    =
    \operatorname{artanh}\!\left(\frac{i}{\sqrt{s}}\right)
    =
    i\arctan\!\left(\frac{1}{\sqrt{s}}\right),
    \qquad (s>0).
    \label{eq:btz_arccoth_positive_axis}
\end{equation}
Substitution into \eqref{eq:btz_I_input} then gives
\begin{equation}
    \mathcal{I}(s)
    =
    \frac{\operatorname{arccoth}(-i\sqrt{s})}{i\sqrt{s}}
    =
    \frac{\arctan(1/\sqrt{s})}{\sqrt{s}}.
    \label{eq:btz_I_positive_axis}
\end{equation}
If the continuation path instead passes below the origin, the continuous argument of \(-x\)
increases from \(0\) to \(+\pi\), giving the conjugate choice
\(E=-i\sqrt{s}\).  Substitution into
\eqref{eq:btz_I_input} again gives \eqref{eq:btz_I_positive_axis}, so the two auxiliary continuations yield the same Abel input on the positive real axis.

Replacing the integration variable \(x\) with \(s\) on the positive real axis, the Abel transform becomes
\begin{equation}
    \mathcal{J}(W)
    =
    \int_{W}^{\infty}
    \frac{\arctan(1/\sqrt{s})}{\sqrt{s}\sqrt{s-W}}\,ds.
    \label{eq:btz_mathcalJ_W}
\end{equation}
By setting \(s=W\cosh^2v\), one can obtain
\begin{equation}
    \mathcal{J}(W)
    =
    2\int_{0}^{\infty}
    \arctan\!\left(\frac{1}{\sqrt{W}\cosh v}\right)dv
    =\pi\,\operatorname{arsinh}\!\Bigl(\frac{1}{\sqrt{W}}\Bigr),
    \qquad (W>0).
    \label{eq:btz_mathcalJ_result_W}
\end{equation}
Applying the Abel inversion formula now gives
\begin{equation}
    H(W)=-\frac{1}{\pi}\frac{d\mathcal{J}}{dW}
    =\frac{1}{2W\sqrt{1+W}}
    ,
    \qquad (W>0).
    \label{eq:btz_H_final}
\end{equation}
This expression defines an analytic branch away from the pole at \(W=0\) and
the square-root branch point at \(W=-1\).  Its value on the
physical branch \(W<-1\) can be obtained by analytic continuation.

The remaining reconstruction is immediate:
\begin{equation}
    z^{-1}(W)
    =
    \int W H(W)\,dW + C
    =
    \frac{1}{2}\int \frac{dW}{\sqrt{1+W}} + C
    =
    \sqrt{1+W}+C.
    \label{eq:btz_zinv_step}
\end{equation}
{As discussed in the general formalism, we fix \(C\) by imposing the horizon condition \(f(1)=0\). Since \(W=f(z)/z^2\), the horizon corresponds to \(W=0\). Thus \(z^{-1}(0)=1\), which fixes \(C=0\).} With this choice,
\begin{equation}
    z^{-1}(W)=\sqrt{1+W},
    \qquad
    W=z^{-2}-1.
    \label{eq:BTZconstC}
\end{equation}
Using \(W=f(z)/z^2\), we finally obtain
\begin{equation}
    f(z)=1-z^{2}.
\end{equation}
This analytic BTZ example provides an exact consistency check of the
Abel reconstruction based on the complex-coordinate prescription.

\paragraph{Numerical SAdS\(_4\) reconstruction from forward data.}

We now test the inverse construction in the SAdS\(_4\) background.  Unlike the
BTZ case, the TEE is not available in closed analytic form, so the input function is
obtained numerically from the forward method of Sec.~\ref{subsec:forward_method}.
We first trace the selected upper vacuum-connected branch of complex turning points
\(z_t\).  The branch is initialized at the UV end, where
\(\Delta t\to0\), \(z_t\to0\), and \(|E|\to\infty\), and is then followed by
continuous continuation subject to the real-time condition
\begin{equation}
    \mathrm{Im}\,\Delta t(z_t)=0 .
\end{equation}
Along this branch we compute the boundary time width \(\Delta t\), the
regularized TEE \(S_{\rm reg}\), the conserved quantity \(E\), and the
on-shell lower endpoint
\begin{equation}
    W_t(\lambda) \equiv -E(\lambda)^2 .
\end{equation}
Here ``on-shell'' means evaluated on the traced extremal-surface branch.  In
the forward benchmark \(W_t=W(z_t)\), but the Abel inversion step uses only the
endpoint value \(W_t=-E^2\) supplied by the entropy data on the selected
branch.  We
use this distinction below when differentiating along the sampled path.

To reconstruct the bulk information accessible from the time-width data, we define
\begin{equation}
    \mathcal{I}(W_t) \equiv \frac{\Delta t(W_t)}{2E(W_t)} .
    \label{eq:ads4_I_definition}
\end{equation}
The factor of \(1/2\) appears because the integral equation is written for the
one-sided radial integral, whereas \(\Delta t\) is the total boundary time
separation.  From the sampled \(\Delta t(E)\) data we then apply Abel
inversion to recover the Abel density \(H(W)\).  In the numerical
implementation, the derivative \(d\mathcal{I}/dW\) entering the inversion formula is
evaluated along the traced endpoint curve \(W_t(\lambda)\),
\begin{equation}
    \frac{d\mathcal{I}}{dW}
    =
    \frac{d\mathcal{I}/d\lambda}{dW_t/d\lambda},
    \label{eq:dIdW_ads4_lambda}
\end{equation}
where both \(\mathcal{I}(\lambda)\) and \(W_t(\lambda)\) are represented by smooth
interpolants along the branch.  The parameter \(\lambda\) is only a smooth
arclength-type tracing coordinate for the curve \(W_t(\lambda)\); the ratio in
\eqref{eq:dIdW_ads4_lambda} is invariant under smooth reparametrizations.
In this way, Abel inversion uses the boundary data sampled along the forward
branch to reconstruct the density \(H(W)\).  No additional continuation in the
\(x\)-plane is used in this numerical step; the required square-root sheets are
followed along the sampled endpoint curve on the selected branch.

A reconstruction in terms of \(G_3(W)=z^{-3}(W)\), as defined by the \(d=3\)
specialization of Eq.~\eqref{eq:Gd_def_main}, is numerically ill-conditioned
near the AdS boundary, since \(G_3(W)\) diverges as
\(z\to0\).  We therefore introduce the UV-regular variable
\begin{equation}
    u = W^{-1/4},
    \label{eq:u_definition_ads4}
\end{equation}
for which \(u\sim z\) near the boundary.  We correspondingly define the
UV-regularized Abel density
\begin{equation}
    \widehat H(u)=\frac{H(W(u))}{u^5}.
    \label{eq:Hhat_definition_ads4}
\end{equation}
Using Eq.~\eqref{eq:H_def_first} for \(d=3\), together with
\(W=u^{-4}\), the inverse problem is then reformulated as a first-order
equation for the bulk path itself,
\begin{equation}
    \frac{dz}{du}
    =
    4\left(\frac{z}{u}\right)^4 \widehat H(u),
    \label{eq:z_u_equation_ads4}
\end{equation}
For the numerical implementation we use the tracing parameter
\(\lambda\),
\begin{equation}
    \frac{dz}{d\lambda}
    =
    4\left(\frac{z}{u}\right)^4
    \widehat H(\lambda)\frac{du}{d\lambda}.
    \label{eq:z_lambda_equation_ads4}
\end{equation}
This is the equation solved numerically in the SAdS\(_4\) reconstruction, where the Abel density is determined by the
TEE data. {In this \(u\)-regularized formulation, the residual integration freedom can be related to the near-boundary initial data for \eqref{eq:z_u_equation_ads4}.}

Since \(W=u^{-4}\), the blackening factor along a reconstructed profile is
obtained from
\begin{equation}
    f(z)
    =
    z^4W
    =
    \left(\frac{z}{u}\right)^4 .
    \label{eq:f_from_u_ads4}
\end{equation}

To distinguish the information encoded in the input function from the residual
freedom, we examine the near-boundary behavior of the original input
\(\mathcal I\), from which the lower-order UV coefficients can be extracted.

Consider the general local expansion
\begin{equation}
    z(u)=u+a_2u^2+a_3u^3+a_4u^4+\cdots .
    \label{eq:ads4_blind_order_anchor}
\end{equation}
Using Eq.~\eqref{eq:f_from_u_ads4}, the corresponding general expansion of the
blackening factor is
\begin{equation}
    f(z)
    =
    1+4a_2z
    +\left(2a_2^2+4a_3\right)z^2
    +4a_4z^3
    +O(z^4).
    \label{eq:ads4_uv_expansion_structure}
\end{equation}

Substituting Eq.~\eqref{eq:ads4_blind_order_anchor} into
Eq.~\eqref{eq:z_u_equation_ads4} gives
\begin{equation}
    \widehat H(u)
    =
    \frac14-\frac{a_2}{2}u
    +\frac{2a_2^2-a_3}{4}u^2
    +0\cdot u^3+O(u^4).
    \label{eq:Hhat_uv_expansion_ads4}
\end{equation}
The leading value of the UV-regularized Abel density,
\(\widehat H(0)=1/4\), follows from \(z/u\to1\), while the
\(u^3\) coefficient cancels identically.

The corresponding boundary expansion of the input function
\(\mathcal I(u)\equiv\mathcal I(W_t(u))\) is therefore
\begin{equation}
    \mathcal I(u)
    =
    \frac14 B\!\left(\frac12,\frac34\right)u^3
    -\frac{a_2}{2}B\!\left(\frac12,1\right)u^4
    +\left(\frac{a_2^2}{2}-\frac{a_3}{4}\right)
     B\!\left(\frac12,\frac54\right)u^5
    +O(u^6),
    \label{eq:I_uv_readout_ads4}
\end{equation}
where \(B\) denotes the Euler beta function.  Hence the \(u^4\) coefficient
of the input function determines \(a_2\), while its \(u^5\) coefficient determines
\(a_3\) once \(a_2\) is known.

For the SAdS\(_4\) data, fitting the \(\mathcal I\)-level UV readout gives
\begin{equation}
    a_2=5.187\times10^{-9},\qquad
    a_3=-8.168\times10^{-10},
\end{equation}
so both coefficients are consistent with zero at the level of the numerical
data.

The origin of the remaining freedom can be seen directly from
Eq.~\eqref{eq:z_u_equation_ads4}, which may be written as
\begin{equation}
    \frac{d}{du}z^{-3}(u)
    =
    -\frac{12}{u^4}\widehat H(u),
    \qquad
    z^{-3}(u)
    =
    -12\int^u \frac{\widehat H(v)}{v^4}\,dv+C .
    \label{eq:ads4_radial_integration_constant}
\end{equation}
Thus, once the Abel density is fixed, the radial profile contains only one
independent integration constant \(C\).  The near-boundary expansion
\eqref{eq:ads4_blind_order_anchor} gives
\begin{equation}
\begin{aligned}
    z^{-3}(u)
    ={}&
    u^{-3}
    -3a_2u^{-2}
    +\left(6a_2^2-3a_3\right)u^{-1}
    \\
    &+\left(-10a_2^3+12a_2a_3-3a_4\right)
    +O(u).
\end{aligned}
    \label{eq:ads4_zinverse_uv_expansion}
\end{equation}
The coefficients \(a_2\) and \(a_3\) are determined by the near-boundary
behavior of \(\mathcal I\), while \(a_4\) parametrizes the remaining
\(O(1)\) integration constant.  More generally, an omitted term
\(a_nu_0^n\) with \(n\geq5\) changes \(z^{-3}(u_0)\) only by
\(O(u_0^{\,n-4})\), which is higher order than the \(O(1)\) contribution of
\(C\) and vanishes as \(u_0\to0\). Therefore, we initialize
the radial equation as
\begin{equation}
    z(u_0)
    =
    u_0+a_2u_0^2+a_3u_0^3+a_4u_0^4+O(u_0^5).
    \label{eq:ads4_uv_anchor_general}
\end{equation}
In a genuine inverse problem, \(a_4\) may also be supplied by an independent
spacelike-entanglement-entropy reconstruction of the near-boundary expansion
of \(f(z)\)~\cite{Bilson_2011,Jokela:2020auu,Xu_2023}.

Starting from the finite-cutoff profile in
Eq.~\eqref{eq:ads4_uv_anchor_general}, we integrate
Eq.~\eqref{eq:z_lambda_equation_ads4} along the reconstructed branch.  Once the
path \(z(u)\) has been fixed for a trial value of \(a_4\), the corresponding
blackening factor follows from Eq.~\eqref{eq:f_from_u_ads4}.  With this
separation of the data-determined and residual coefficients, the numerical
reconstruction pipeline can be summarized schematically as
\begin{equation}
    \bigl\{\Delta t(\lambda),E(\lambda),W_t(\lambda)\bigr\}
    \;\longrightarrow\;
    \mathcal{I}(W_t)
    \;\longrightarrow\;
    H(W)
    \;\longrightarrow\;
    \widehat H(u)
    \;\longrightarrow\;
    z(u)
    \;\longrightarrow\;
    f(z).
    \label{eq:ads4_reconstruction_pipeline}
\end{equation}
The near-boundary behavior of \(\mathcal I\) fixes \(a_2\) and \(a_3\), while
the last two arrows still depend on the trial value of \(a_4\).

\begin{figure}[!htbp]
    \centering
    \includegraphics[width=0.85\textwidth]{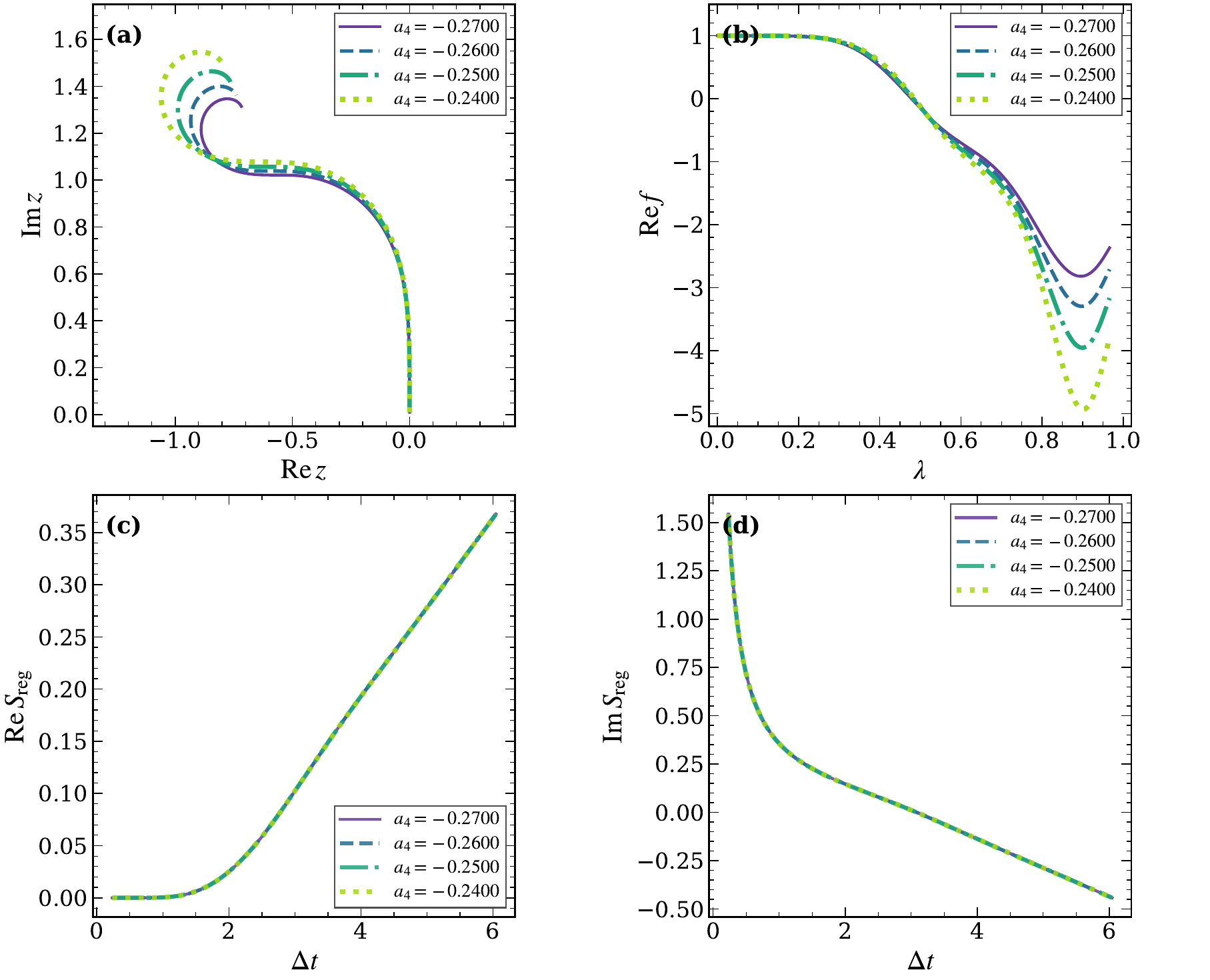}
\caption{Representative ambiguity associated with the residual quartic
(\(u^4\)) UV coefficient \(a_4\) in the SAdS\(_4\) inverse reconstruction.
All reconstructions use the same time-width data \(\Delta t(E)\), reconstructed
Abel density \(H(W)\), and data-determined coefficients \(a_2\) and \(a_3\);
varying only \(a_4\) produces distinct complex bulk paths and blackening
functions. Nevertheless, forward recomputation for each representative, with
\(S_{\rm reg}=A_{\rm reg}/4\), yields overlapping
\(\mathrm{Re}\,S_{\rm reg}(\Delta t)\) and
\(\mathrm{Im}\,S_{\rm reg}(\Delta t)\) curves. The
entropy data therefore do not resolve the remaining blackening-function
freedom.}
    \label{fig:ads4_different_a_same_entropy}
\end{figure}

To make this point explicit, we repeat the final step of the inverse
reconstruction with several choices of the residual quartic (\(u^4\)) UV
coefficient \(a_4\), while keeping the time-width data \(\Delta t(E)\), the
data-determined coefficients \(a_2,a_3\), and the reconstructed Abel density
\(H(W)\) fixed. As illustrated in
Fig.~\ref{fig:ads4_different_a_same_entropy}, these choices yield distinct
reconstructed complex bulk paths \(z(u)\) and, through
Eq.~\eqref{eq:f_from_u_ads4}, different blackening functions. Despite these
differences, each representative reproduces the same entropy data: recomputing
the forward entropy yields overlapping \(S_{\rm reg}(\Delta t)\) curves. Thus, neither the time-width nor the entropy data uniquely
fix the remaining representative freedom, highlighting the need for an
additional profile-fixing anchor when comparing the inverse reconstruction with
a specific benchmark geometry.

To fix this remaining freedom, for each trial \(a_4\), with \(a_2\) and \(a_3\)
held at the values read from \(\mathcal I\), we reconstruct \(z_{a_4}(u)\) and
\(f_{a_4}=(z_{a_4}/u)^4\).  We then apply the Adaptive Antoulas--Anderson (AAA) rational approximation~\cite{Nakatsukasa:2018AAA}
directly to the samples \(\{(z_j,f_{a_4}(z_j))\}\) on the corresponding
complex path.

The AAA algorithm is an adaptive method for constructing an accurate rational approximation of a function from sampled data.  For each trial \(a_4\), we denote by
\(F_m^{(a_4)}(z)\) the approximant constructed from these samples using \(m\)
adaptively selected support points.  It is written in barycentric form as
\begin{equation}
    F_m^{(a_4)}(z)
    =
    \frac{
    \sum_{k=1}^{m} w_k f_{a_4}(\zeta_k)/(z-\zeta_k)
    }{
    \sum_{k=1}^{m} w_k/(z-\zeta_k)
    },
    \label{eq:aaa_barycentric_form}
\end{equation}
where the support points \(\zeta_k\) are selected from the available sample
points and the weights \(w_k\) are determined by a linear least-squares
problem on the remaining samples.  The support points are chosen adaptively:
after each rational fit, the next support point is placed where the current
approximation has the largest residual on the sampled path.  This greedy
selection lets the rational function concentrate its degrees of freedom near
the most difficult parts of the data.

\begin{figure}[!htbp]
	\centering
	\includegraphics[width=0.85\textwidth]{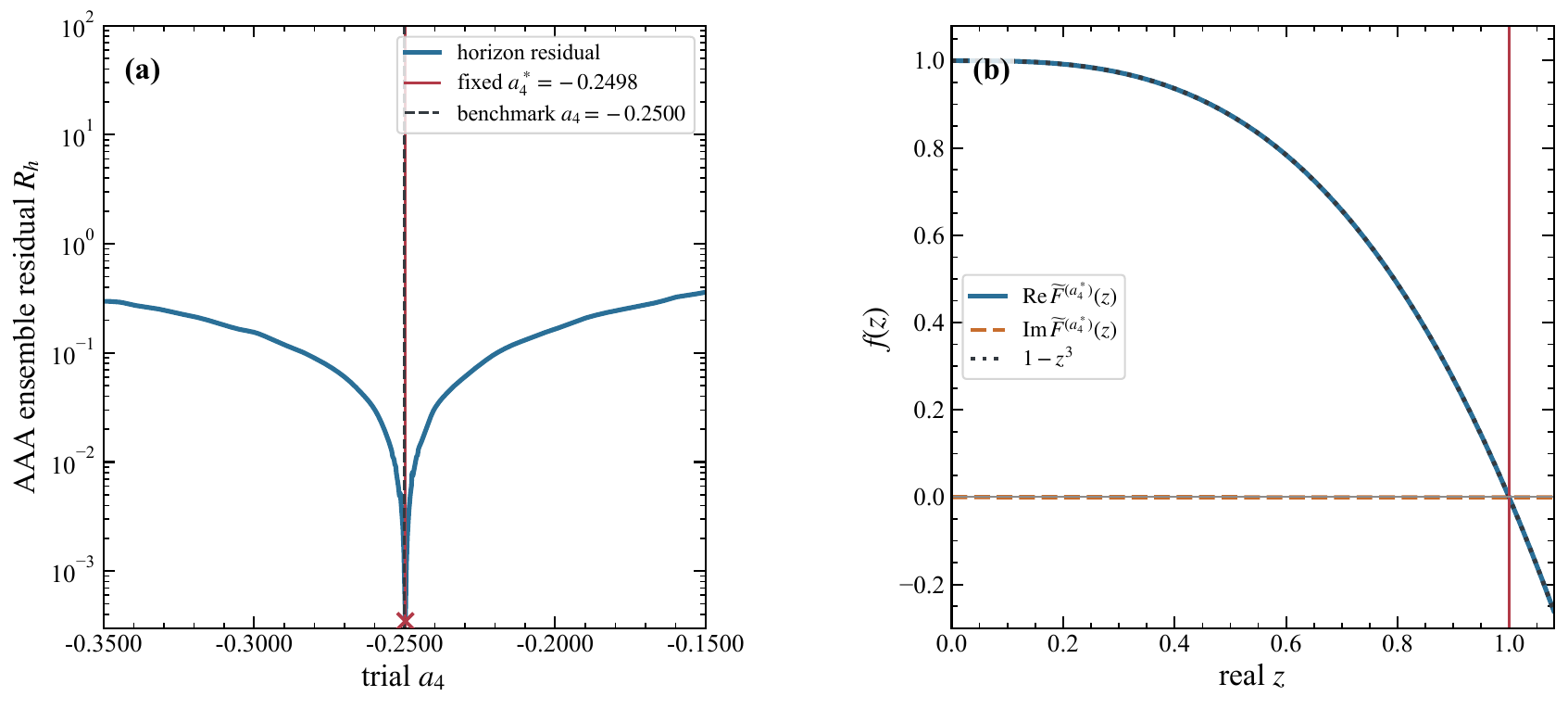}
	\caption{Horizon fixing of the quartic (\(u^4\)) UV coefficient in the
        SAdS\(_4\) reconstruction.  The lower coefficients \(a_2\) and \(a_3\)
        are first read from the near-boundary behavior of the input function
		\(\mathcal I\).  Panel (a) scans the horizon residual over the remaining
		coefficient \(a_4\); the dashed vertical line marks the benchmark value
		\(a_4=-1/4\).  Panel (b) restricts the ensemble continuation at the
		selected \(a_4^*\) to the real axis and compares its real and imaginary
		parts with the SAdS\(_4\) form \(1-z^3\).}
	\label{fig:ads4_aaa_horizon_fixing}
\end{figure}

We then use the horizon position to specify $a_4$. Condition \(f(1)=0\) is applied only after
the continuation has been constructed.  We form an AAA ensemble by varying the
maximum number of support points, the stopping tolerance, and the fraction of
the reconstructed path used for fitting.  For the accepted fits entering
Fig.~\ref{fig:ads4_aaa_horizon_fixing}, the selected number of support points
lies in the range \(7\leq m\leq18\).  We denote the median continuation over
the accepted ensemble by \(\widetilde F^{(a_4)}(z)\), and define
\begin{equation}
    R_h(a_4)=\left|\widetilde F^{(a_4)}(1)\right|.
\end{equation}

As shown in Fig.~\ref{fig:ads4_aaa_horizon_fixing}(a), the \(a_4\) scan selects
\begin{equation}
    a_4^*=-0.2498,\qquad R_h=3.476\times10^{-4}.
    \label{eq:ads4_a4_scan_result}
\end{equation}
Since the \(a_4\) term gives
\(f(z)=1+4a_4 z^3+\cdots\), this corresponds to
\(4a_4^*=-0.9990\), close to the SAdS\(_4\) coefficient \(-1\).  The result
supports the expected separation: the low-order UV coefficients are fixed by
the near-boundary structure of \(\mathcal I\), while the quartic (\(u^4\)) UV
coefficient is the remaining freedom selected by the horizon condition.

\begin{figure}[!htbp]
    \centering
    \includegraphics[width=0.85\textwidth]{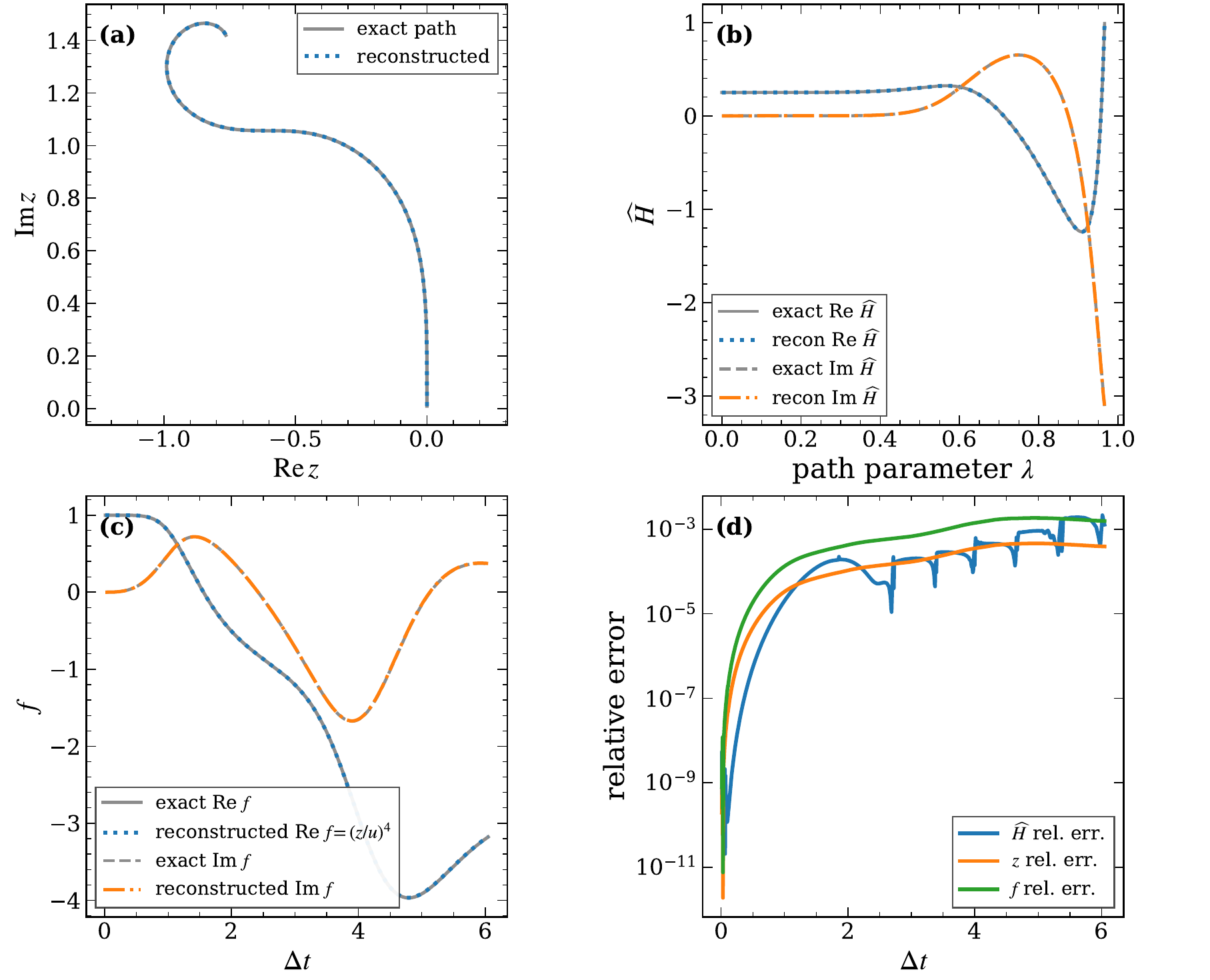}
    \caption{Numerical SAdS\(_4\) reconstruction after fixing the bulk path
    by the selected UV coefficient. Panel (a) compares the selected complex
    trajectory with the exact turning-point path. Panel (b) shows the Abel reconstruction of
    \(\widehat{H}(\lambda)\) along the traced branch. Panel
    (c) compares the real and imaginary parts of the reconstructed blackening
    factor with the exact SAdS\(_4\) result. Panel (d) shows the relative
    reconstruction errors as functions of \(\Delta t\).}
    \label{fig:ads4_reconstruction_summary}
\end{figure}

As summarized in Fig.~\ref{fig:ads4_reconstruction_summary}, the reconstructed
UV-regularized Abel density
\(\widehat H(\lambda)\) agrees well with the exact benchmark along the traced
branch truncated before the critical point.  This directly checks that Abel
inversion extracts the expected density from the forward data on this branch.
After the bulk path is fixed by the UV
coefficient, the corresponding \(z(u)\) and
blackening factor \(f(z)\) also agree with the SAdS\(_4\) benchmark. The relative-error plot shows that the reconstruction remains well controlled
over most of the traced branch.  The largest deviations occur near the endpoint,
where the trajectory approaches the internal critical point and the inverse map
becomes increasingly sensitive to numerical error.

\subsection{Applications of the complex-coordinate method}
\label{sec:charged_applications}

We now apply the complex-coordinate method of Sec.~\ref{sec:3} to two charged
backgrounds.
{The forward computation traces a selected vacuum-connected complex branch to obtain \(S_{\rm reg}(\Delta t)\), and the Abel inversion is subsequently applied to these branch data. We first test the one-function reconstruction in the RN geometry, and then investigate the Gubser--Rocha model where a nontrivial spatial warp factor \(h(z)\) is present.}

\subsubsection{Reissner--Nordstr\"om black hole}
\label{subsec:RN_application}

\begin{figure}[ht]
    \centering
    \includegraphics[width=0.85\textwidth]{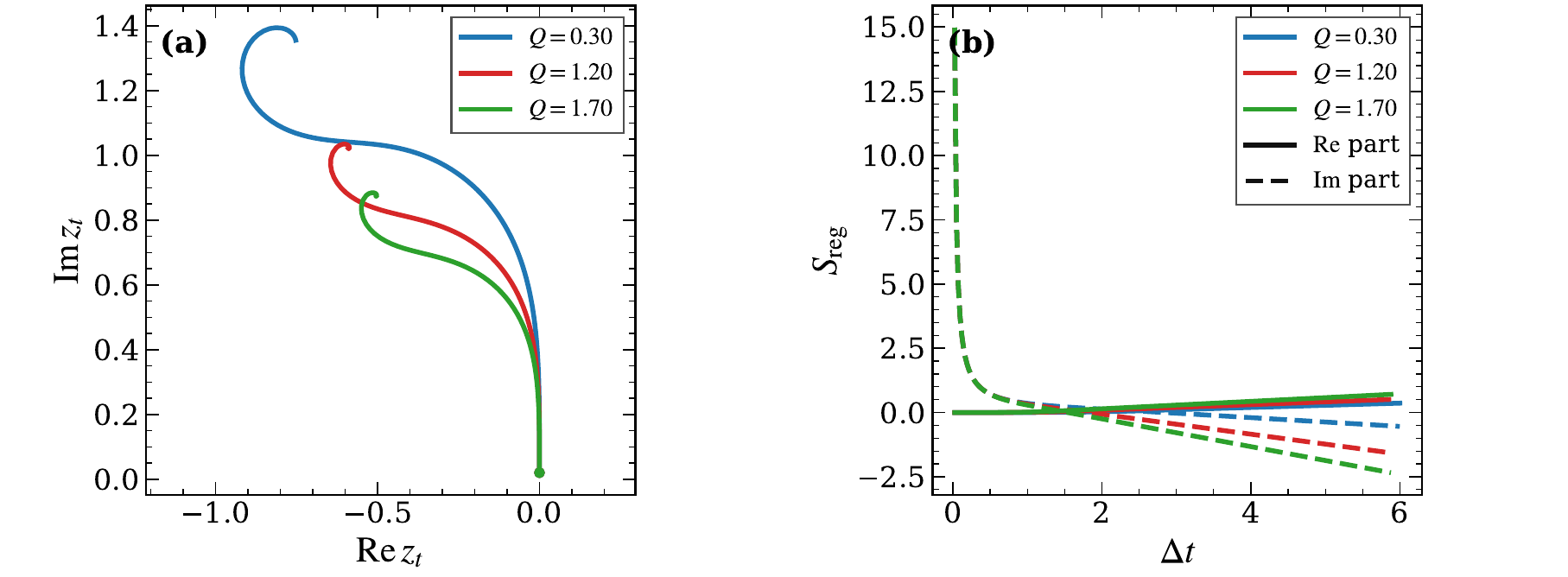}
    \caption{Forward results of the complex-coordinate method in the RN background.
    Left: turning-point trajectories in the complex plane.
    Right: real and imaginary parts of the regularized TEE.}
    \label{fig:RN_results_main}
\end{figure}

We first consider the planar RN black hole in
AdS\(_4\), with the blackening factor
\begin{equation}
    f_{\mathrm{RN}}(z)
    =
    1-(1+Q^2)z^3+Q^2 z^4,
    \label{eq:f_RN_application}
\end{equation}
where the outer horizon is fixed at \(z=1\).  The allowed charge range is
\(0\le Q\le \sqrt{3}\).  For each \(Q\), the branch is initialized at
the vacuum-connected UV endpoint and then traced with
\(\mathrm{Im}\,\Delta t=0\), oriented so that \(\Delta t\) increases along the
branch.  This gives the turning-point trajectory \(z_t\) and the regularized
entropy \(S_{\rm reg}(\Delta t)\).  The resulting branch trajectories and
entropy curves are displayed in Fig.~\ref{fig:RN_results_main}.  The charge
deformation changes both the trajectory and the entropy profile, while the
forward construction remains smooth on the chosen branch.

The large-\(\Delta t\) limit is controlled by the critical point of
\(W(z)=f_{\rm RN}(z)/z^4\).  As the endpoint \(-E^2=W(z_t)\) approaches a
critical value \(W(z_c)\) with \(dW/dz|_{z_c}=0\), the inverse map develops
\(dz/dW\sim (W-W_c)^{-1/2}\).  The Abel density therefore has a square-root
endpoint singularity, and the main contribution to the time and entropy
integrals comes from a small neighborhood of \(z_c\).  This fixes the
late-time linear slope.  For RN,
\begin{equation}
    (1+Q^2)z_c^3=4,\qquad
    z_c^{(n)}
    =
    \left(\frac{4}{1+Q^2}\right)^{1/3}e^{2\pi i n/3}.
    \label{eq:RN-zc-roots}
\end{equation}
The branch used in the numerical calculation selects the \(n=1\) root, with
\(\mathrm{Im}\,z_c>0\).  Concretely, this is the root approached continuously by
the traced \(z_t\) trajectory as \(\Delta t\) increases.  Continuing the same
\(E(z_t)\) branch fixes \(E_c\) through
\(E_c^2=-W(z_c)=3/z_c^4-Q^2\).  The late-time slope is
\begin{equation}
    \left.
    \frac{dS_T}{d\Delta t}
    \right|_{\Delta t\to\infty}^{\mathrm{RN}}
    =
    -\frac{E_c}{4}.
    \label{eq:RN-slope-final}
\end{equation}
Using this critical-point formula for the three representative RN backgrounds
gives the approximate values
\begin{equation}
\begin{aligned}
    \left.
    \frac{dS_{\rm reg}}{d\Delta t}
    \right|_{\Delta t\to\infty}^{Q=0.3}
    &\simeq 0.08390 - 0.1710\,i,\\
    \left.
    \frac{dS_{\rm reg}}{d\Delta t}
    \right|_{\Delta t\to\infty}^{Q=1.2}
    &\simeq 0.1084 - 0.3876\,i,\\
    \left.
    \frac{dS_{\rm reg}}{d\Delta t}
    \right|_{\Delta t\to\infty}^{Q=1.7}
    &\simeq 0.1448 - 0.5403\,i.
\end{aligned}
\label{eq:RN_large_t_slope_values}
\end{equation}

\begin{figure}[ht]
    \centering
    \includegraphics[width=0.62\textwidth]{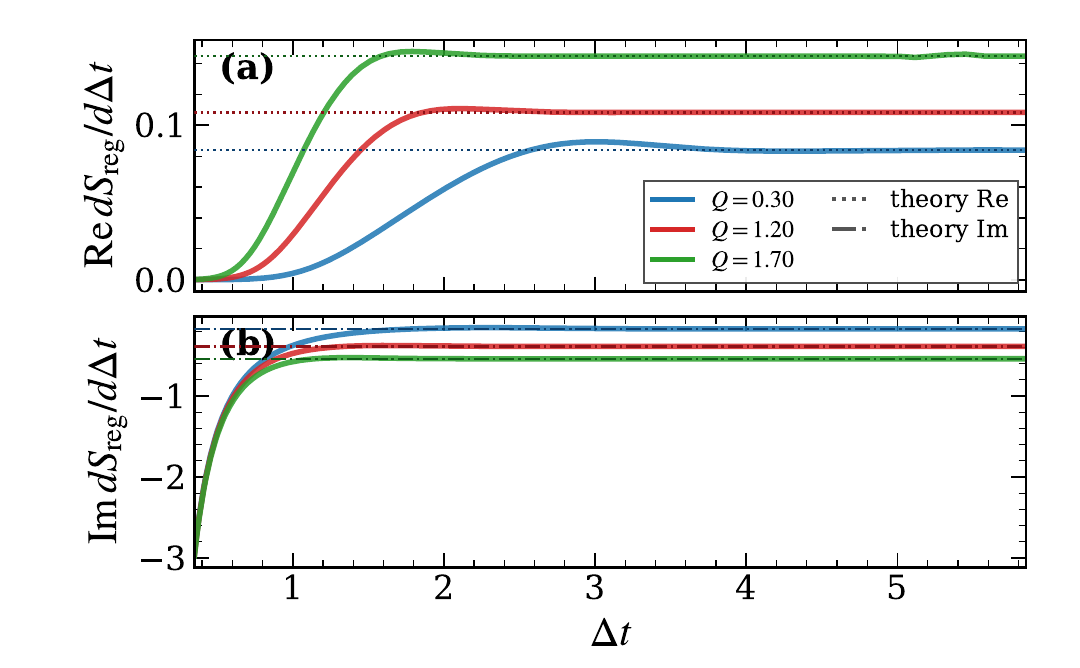}
    \caption{Derivative \(dS_{\rm reg}/d\Delta t\) in the RN background.
The thick colored curves show the numerical derivatives for the three
charges. The thin horizontal lines denote the late-time values in
Eq.~\eqref{eq:RN_large_t_slope_values}.}
    \label{fig:RN_derivative}
\end{figure}

As shown in Fig.~\ref{fig:RN_derivative}, the numerical derivatives approach
these critical-point values in the late-time region. This provides a non-trivial check of the forward numerical algorithm.

\paragraph{Numerical reconstruction in the RN background.}

Since \(h(z)=1\), the inverse problem is the same one-function problem as in
the SAdS\(_4\) case, with a charged blackening factor and a different
near-boundary behavior.  For reference, inverting
\(u=W^{-1/4}=z f_{\rm RN}(z)^{-1/4}\) for
\eqref{eq:f_RN_application} gives the exact RN benchmark
\begin{equation}
    z(u_0)
    =
    u_0
    -\frac{1+Q^2}{4}u_0^4
    +\frac{Q^2}{4}u_0^5
    +O(u_0^7) .
    \label{eq:RN_z_u_expansion}
\end{equation}
Thus,
\[
a_4^{\rm exact}=-\frac{1+Q^2}{4},
\qquad
a_5^{\rm exact}=\frac{Q^2}{4}.
\]
For \(Q=1.2\), these give
\(a_4^{\rm exact}=-0.61\) and
\(a_5^{\rm exact}=0.36\).

In the numerical reconstruction, the near-boundary
Abel data first determine
\(a_2=6.750\times10^{-8}\) and
\(a_3=2.255\times10^{-6}\). The coefficient multiplying \(u^7\) in Eq.~\eqref{eq:I_uv_readout_ads4} imposes the constraint between $a_4$ and $a_5$ by%
\begin{equation}
\bigl[u^7\bigr]\mathcal{I}(u)
=
\frac14 B\!\left(\frac12,\frac74\right)
\left(
a_5-5a_2^4+10a_2^2a_3
-4a_2a_4-2a_3^2
\right).
\label{eq:I_u7_uv_constraint}
\end{equation}
We then perform a one-parameter constrained horizon-fixing scan in the
\((a_4,a_5)\) plane: for each trial \(a_4\), the corresponding \(a_5\)
is fixed by this Abel constraint, and the resulting pair is evaluated
using the horizon residual. The scan selects
\((a_4^*,a_5^*)=(-0.6103,0.3595)\).\footnote{As discussed above Eq.~\eqref{eq:ads4_uv_anchor_general}, terms beyond \(u_0^4\) vanish as \(\epsilon\to0\).
At our finite cutoff, however, omitting \(a_5\) in the RN example can produce a relative error of
up to \(4\%\) in the reconstructed \(f\) near the IR endpoint. We therefore
retain the available \(a_5\) information.}
\begin{figure}[!htbp]
    \centering
    \includegraphics[width=0.85\textwidth]{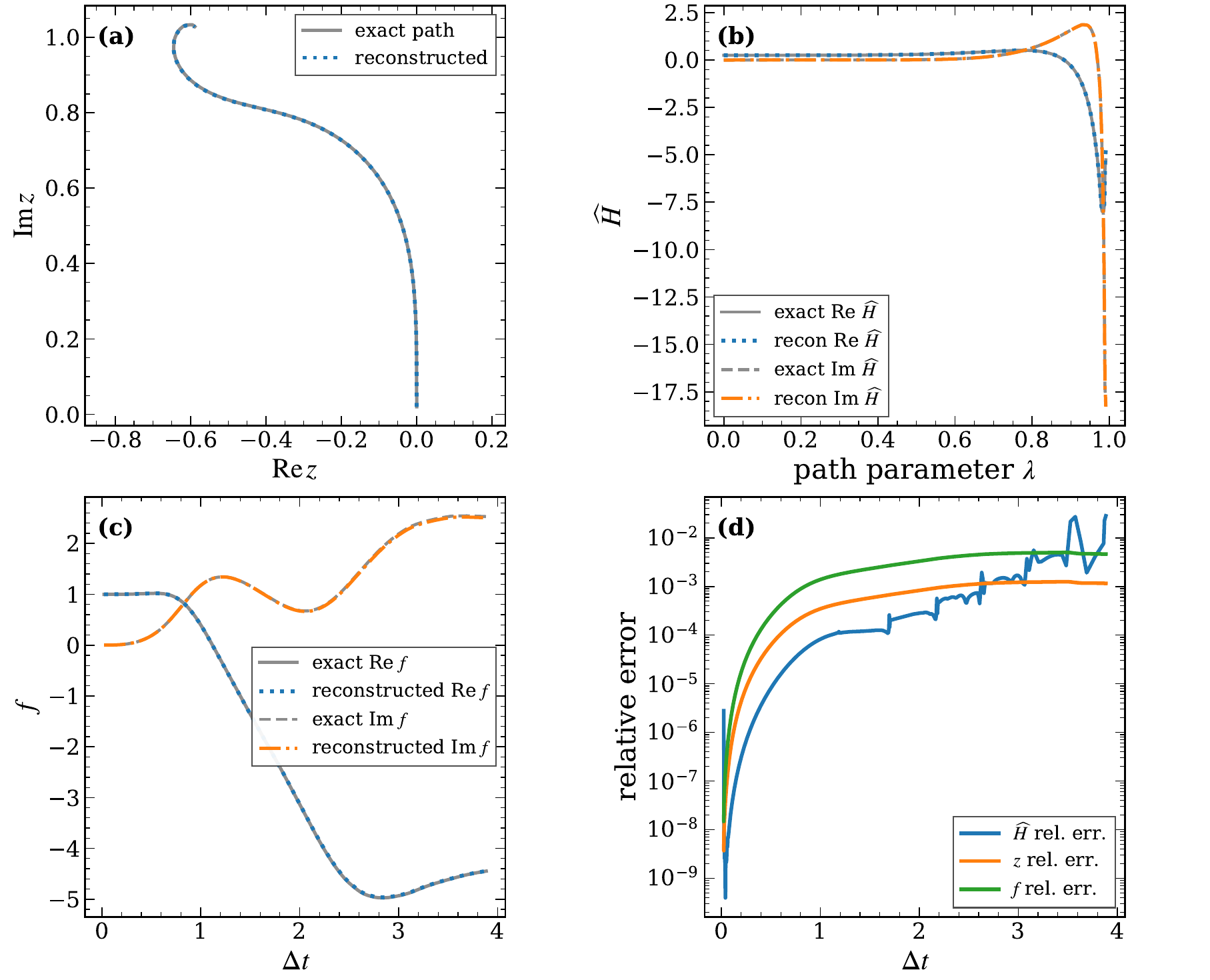}
    \caption{Numerical reconstruction in the RN background (\(Q=1.2\))
    using the selected UV coefficient.
    Panel (a) compares the reconstructed complex trajectory with the exact
    turning-point path. Panel (b) shows the Abel reconstruction of
    \(\widehat H(\lambda)\).  Panel (c) compares the real and imaginary parts of
    the reconstructed blackening factor with the exact RN result.  Panel (d)
    summarizes the relative reconstruction errors.}
    \label{fig:rn_reconstruction_summary}
\end{figure}

The reconstruction results for the \(Q=1.2\) RN background
are summarized in Fig.~\ref{fig:rn_reconstruction_summary}.  The reconstructed
trajectory, Abel density, and blackening factor are all compared with their exact
benchmarks on the same traced branch, truncated before the critical point.  The
agreement shows that the charged deformation is compatible with the Abel inversion:
after the RN-specific anchor is supplied, the \(u\)-regularized reconstruction
recovers the complex path and the blackening factor.  The largest deviations
occur near the endpoint, where the inverse map becomes sensitive to the nearby
critical point.

\subsubsection{Gubser--Rocha model}
\label{subsec:GR_application}

We next consider the Gubser--Rocha background
\begin{equation}
    ds^2=
    \frac{1}{z^2}
    \left[
        -f(z)\,dt^2+\frac{dz^2}{f(z)}+h(z)\left(dx^2+dy^2\right)
    \right],
    \label{eq:GR-metric}
\end{equation}
with the blackening factor and a nontrivial spatial warp factor:
\begin{equation}
    f(z)
    =
    (1-z)\,
    \frac{
        1+(1+3Q)z+\left(1+3Q(1+Q)\right)z^2
    }{
        (1+Qz)^{3/2}
    },
    \qquad
    h(z)=(1+Qz)^{3/2}.
    \label{eq:GR-fh}
\end{equation}
The numerical plots are labeled by the equivalent parameter
\(\mu^2=3Q(1+Q)\).
The fractional power is defined by the branch that is real and positive at the
AdS boundary, \((1+Qz)^{3/2}=1\) at \(z=0\), and is then continued along the
chosen contour.  The contours used below do not cross the branch point
\(z=-1/Q\).

\begin{figure}[ht]
    \centering
    \includegraphics[width=0.85\textwidth]{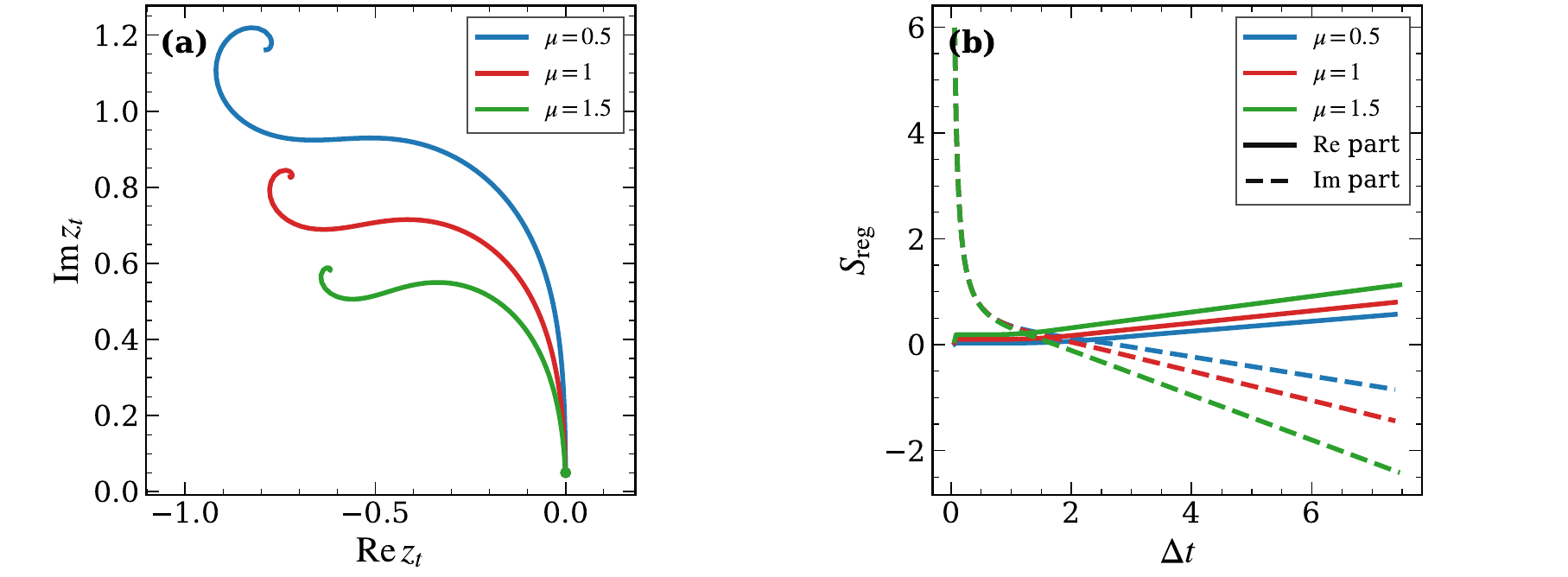}
    \caption{Forward results of the complex-coordinate method in the Gubser--Rocha background.
    Left: turning-point trajectories in the complex plane.
    Right: real and imaginary parts of the regularized TEE.}
    \label{fig:GR_results_main}
\end{figure}

The forward branch is selected as before by starting from the vacuum-connected
UV endpoint and imposing \(\mathrm{Im}\,\Delta t=0\), with the orientation chosen
so that \(\Delta t\) increases along the trajectory.
The UV subtraction is still the standard area-law subtraction.  The resulting
branch trajectories and entropy curves are shown in Fig.~\ref{fig:GR_results_main}.

We introduce
\begin{equation}
W(z)\equiv \frac{f(z)h(z)}{z^4}
\label{eq:GR_W_def}
\end{equation}
as the transformed variable on the selected Gubser--Rocha branch. In the Abel inversion
step the endpoint coordinate is still supplied by the boundary
data through \(W_t=-E^2\), as in the general construction above.  The
critical-point condition
\(dW/dz|_{z_c}=0\) gives
\begin{equation}
    -4
    -9Q z_c
    -6Q^2 z_c^2
    +\left(1+3Q(1+Q)\right)z_c^3
    =0.
    \label{eq:GR-zc-cubic}
\end{equation}
We select the root approached continuously by the traced branch, with
\(\mathrm{Im}\,z_c>0\).  Continuing the same \(E(z_t)\) branch fixes
\(E_c^2=-W(z_c)=-f(z_c)h(z_c)/z_c^4\).  The late-time slope is
\begin{equation}
    \left.
    \frac{dS_T}{d\Delta t}
    \right|_{\Delta t\to\infty}^{\mathrm{GR}}
    =
    -\frac{E_c}{4}.
    \label{eq:GR-slope-final}
\end{equation}
Using this critical-point formula for the three representative values used in
the plots gives the approximate values
\begin{equation}
\begin{aligned}
    \left.
    \frac{dS_{\rm reg}}{d\Delta t}
    \right|_{\Delta t\to\infty}^{\mu=0.5}
    &\simeq 0.09420 - 0.1826\,i,\\
    \left.
    \frac{dS_{\rm reg}}{d\Delta t}
    \right|_{\Delta t\to\infty}^{\mu=1.0}
    &\simeq 0.1163 - 0.2770\,i,\\
    \left.
    \frac{dS_{\rm reg}}{d\Delta t}
    \right|_{\Delta t\to\infty}^{\mu=1.5}
    &\simeq 0.1486 - 0.4229\,i.
\end{aligned}
\label{eq:GR_large_t_slope_values}
\end{equation}

\begin{figure}[ht]
    \centering
    \includegraphics[width=0.62\textwidth]{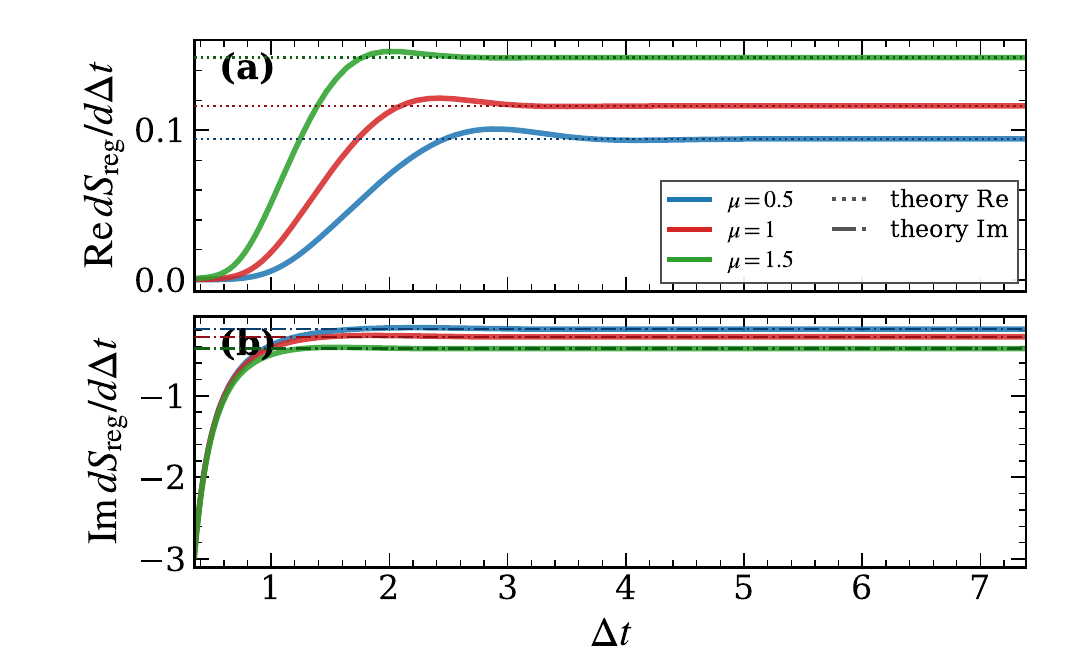}
    \caption{Derivative \(dS_{\rm reg}/d\Delta t\) in the Gubser--Rocha background.
    The thin lines denote the late-time values in
    Eq.~\eqref{eq:GR_large_t_slope_values}.}
    \label{fig:GR_derivative}
\end{figure}
The numerical derivatives in Fig.~\ref{fig:GR_derivative} approach these
critical-point values, confirming the expected late-time behavior.

\paragraph{Abel inversion for the Gubser--Rocha model}

For the Gubser--Rocha background, half of the boundary time width is
\begin{equation}
    \frac{\Delta t(E)}{2}
    =
    E\int_{\bar C(\epsilon,z_t)}
    \frac{z^2\,dz}{f(z)\sqrt{f(z)h(z)+E^2 z^4}} .
    \label{eq:GR_T_E}
\end{equation}
The turning point \(z_t(E)\) is determined by
\begin{equation}
f(z_t)h(z_t)+E^2 z_t^4=0.
\end{equation}

Using the definition \eqref{eq:GR_W_def} to rewrite the integrand on the
selected square-root sheet gives
\begin{equation}
    \frac{\Delta t(E)}{2E}
    =
    \int_{\bar C(\epsilon,z_t)}
    \frac{dz}{f(z)\sqrt{W(z)+E^2}}.
\end{equation}
To change variables from \(z\) to \(W\), we use the same
assumption as stated above Eq.~\eqref{eq:Deltat_over_2E_abel}.
The numerical Abel inversion
below is performed on the traced branch before it reaches the corresponding critical point. The Abel density is defined as
\begin{equation}
H(W)\equiv -\frac{1}{f(z(W))}\frac{dz}{dW}.
\label{eq:GR_H_def}
\end{equation}
Since \(W(z_t)=-E^2\), the width obeys an Abel integral equation
\begin{equation}
    \mathcal{I}(-E^2)\equiv \frac{\Delta t(E)}{2E}
    =
    \int_{-E^2}^{\infty}\frac{H(W)}{\sqrt{W+E^2}}\,dW.
    \label{eq:GR_Abel}
\end{equation}
Equation~\eqref{eq:GR_Abel} has the same form involving the adjoint Abel
operator as Eq.~\eqref{eq:abel_standard_main}.
Here the upper endpoint \(\infty\) denotes the regulated boundary limit
\(W(\epsilon)\to\infty\).  Since the asymptotic AdS normalization gives
\(f(0)h(0)=1\), one has \(W(z)\sim z^{-4}\) near \(z=0\) on the chosen
near-boundary branch.  The integral is understood along the
corresponding path on \(\Gamma\).  The inversion formula gives
\begin{equation}
H(W)
=
-\frac{1}{\pi}\frac{d}{dW}
\int_{W}^{\infty}
\frac{\mathcal{I}(x)}{\sqrt{x-W}}\,dx.
\label{eq:GR_Abel_inv}
\end{equation}
\begin{figure}[!htbp]
    \centering
    \includegraphics[width=0.463\textwidth]{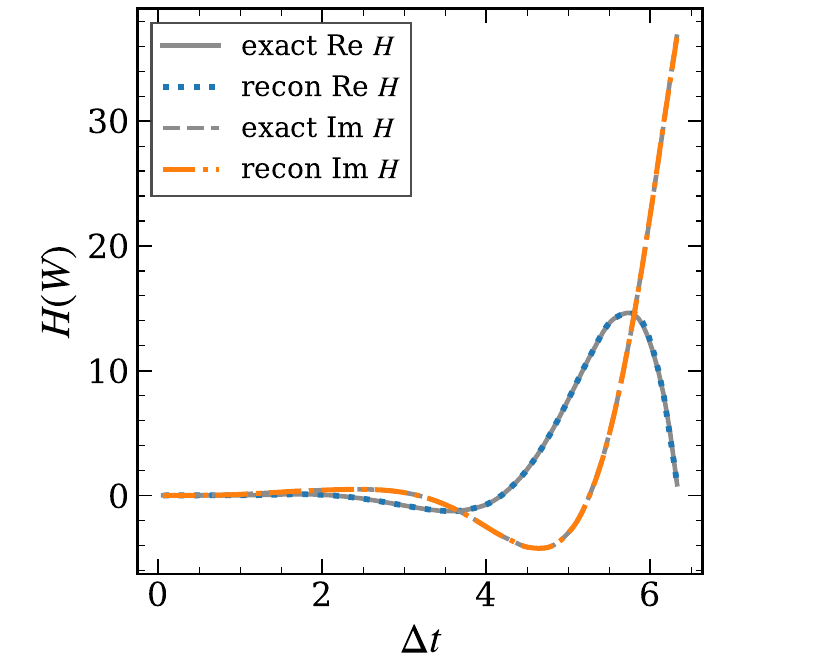}
    \caption{Abel reconstruction of the density \(H(W)\) in the
    Gubser--Rocha model (\(\mu=1.0\)). The horizontal axis is the real boundary time
    \(\Delta t\), used here as a convenient parameter along the branch. The real and imaginary parts of the reconstructed
    Abel density are compared with the exact benchmark evaluated on the same traced
    complex branch.}
    \label{fig:gr_Hw_reconstruction}
\end{figure}

In this numerical test, we apply the Abel inversion to the Gubser--Rocha branch
data obtained from the forward computation at \(\mu=1.0\).  The reconstructed
\(H(W)\) agrees with the exact benchmark over the traced branch, see Fig.~\ref{fig:gr_Hw_reconstruction}.  This confirms
Abel inversion itself, while leaving the metric underdetermination intact: the
strip data fix the density \(H(W)\), but not the two functions \(f\) and \(h\)
separately.
The essential difference from the one-function case is that Abel inversion
returns only the single density \(H(W)\).  Together with the definition of \(W\), it
gives
\begin{equation}
    H(W)=-\frac{1}{f(z(W))}\frac{dz}{dW},
    \qquad
    f(z(W))h(z(W))=Wz(W)^4 .
    \label{eq:GR_metric_constraints}
\end{equation}
These are two relations for the three branch functions
\(z(W)\), \(f(z(W))\), and \(h(z(W))\).  The strip data therefore do not
separate \(f\) and \(h\).  If \(h(z)\) is supplied as extra input, then
\eqref{eq:GR_metric_constraints} gives a first-order equation for \(z(W)\), and
\(f\) follows from \(fh=Wz^4\).  If \(f(z)\) is supplied instead, the first
relation gives the equation for \(z(W)\), and \(h\) follows from the second
relation.  Without such extra input, the metric reconstruction is
underdetermined.

The real and imaginary parts of \eqref{eq:GR_Abel} do not change this counting:
they are two components of the same complex Abel integral equation and
reconstruct the same density \(H(W)\), not a second independent metric
constraint.


\section{Conclusion and Discussion}
\label{sec:conclusion}

We have studied the inverse problem of reconstructing bulk metric functions
from strip-shaped holographic timelike entanglement entropy, developing two
complementary methods with particular emphasis on the dependence of the holographic prescription
and the selected extremal-surface branch.

The first method operates on data associated with the selected CWES
branch: two spacelike branches running from the boundary endpoints to the
future and past singularities, connected by a timelike codimension-two surface
through the bifurcation surface.  The second is a complex-coordinate construction in
which the extremal surface and its square-root sheet are transported through the
complexified bulk.  For the BTZ black hole, both methods yield fully analytic
reconstructions --- the Hamilton--Jacobi relation, integral inversion, and
blackening factor are all obtained in closed form once the appropriate
geometric anchor (singularity endpoint or horizon position) is supplied.  For
backgrounds without closed-form HTEE, we developed a forward numerical method
for the complex-coordinate prescription that traces a vacuum-connected family of
complex turning points while transporting the integration contour continuously.
After validation in SAdS\(_4\)~\cite{Heller_2025}, the procedure was applied to
planar RN--AdS\(_4\) and the Gubser--Rocha background; in both cases, the
numerically extracted late-time entropy growth rates agree with the
corresponding critical-point predictions.

Despite their distinct geometric origins, the two methods share a common
mathematical core: the entropy slope first fixes the conserved quantity~\(E\) via the Hamilton--Jacobi
relation; the time-width relation $\Delta t(E)$ is then reduced to a square-root
transform whose inversion yields an Abel density~\(H(W)\). In the BTZ benchmark
both methods give the same analytic density
\(H(W)=1/(2W\sqrt{1+W})\), although they initially access different domains and
sheets.  This shared structure shows that the inverse problem can sometimes
isolate a common piece of geometric information from the entropy data even
when its bulk realization is prescription dependent. More generally, the
Abel density should be viewed as the geometric information accessible to the
chosen TEE observable on a specified branch; converting it into a definite
radial metric profile requires additional geometric input.

The inverse viewpoint also exposes how analytic continuation enters at
different stages in each method.  The CWES-based reconstruction
continues the boundary input function and extracts \(H(W)\) from the
discontinuity across a branch cut.  The complex-coordinate prescription
instead transports the extremal surface and its square-root sheet through the
complexified bulk.  In the analytic BTZ calculation, a further continuation
extends the physical input to an auxiliary integration domain.  In contrast,
AAA rational approximation acts only after the Abel inversion, interpolating
samples of the reconstructed metric.  Keeping these operations distinct
reveals that analytic continuation enters the TEE problem at several
conceptually different levels: in the boundary data, in the bulk extremal
surfaces and their sheets, and in the subsequent metric reconstruction.  This
layered analytic structure may contain information beyond its immediate role
in the inversion procedure.

These analytic continuations should in turn be distinguished from the
saddle-selection problem. Neither method determines the saddle-selection rule; each acts on a
prescribed smooth branch.  For CWES, weak extremality at the joints and the
ordering of candidate complex areas are upstream selection conditions.  For the
complex-coordinate construction, the turning-point branch, contour class, and
square-root sheet must be fixed beforehand. The entropy curve therefore constrains the geometry conditional on this holographic choice rather than determining the choice itself.

The two prescriptions also differ in their access to the black-hole
interior.\footnote{Alternative approaches reconstruct black-hole interiors
from holographic complexity~\cite{Hashimoto_2021,Xu_2023} or pole-skipping
data~\cite{gmd3-zt39,lgkz-gyl1}; a systematic comparison with the present TEE
framework is left for future work.}
In the selected CWES branch considered here, the extremal surface is embedded
in the real Lorentzian geometry and extends through the horizon into the
singularity. The reconstruction is therefore performed along a real
radial branch and directly probes the interior. By contrast, in the
complex-coordinate prescription, the extremal surface follows a contour in the
complexified radial plane, so the reconstruction directly determines the
geometry along that selected complex path whose late-time behavior is governed by
the critical structure of the map \(z\mapsto W(z)\) rather than by the singularity on the real axis. Recovering the black-hole interior therefore requires a
further analytic continuation from the reconstructed complex path to the real interior branch.

A further identifiability limitation arises when more than one metric
function is unknown.  In the Gubser--Rocha model, where the spatial warp factor
is nontrivial, a single strip observable determines only one functional
combination of \(f(z)\) and \(h(z)\).  This limitation reflects the data content
of the strip geometry rather than the Abel inversion itself; additional
observables --- different strip orientations, alternative region shapes, or
independent probes --- could supply the missing constraints needed to separate
the two metric functions.

Finally, both methods require that the selected branch admits the change of variables
underlying the Abel reduction; for the complex-coordinate construction, the validation of these conditions is nontrivial. In
Appendix~\ref{app:btz_abel_conditions}, we prove analytically that they hold for
BTZ.  We have also checked them numerically for the
SAdS\(_4\), RN--AdS\(_4\), and Gubser--Rocha backgrounds.  Because exact
benchmark metrics are available in these examples, the agreement of the
reconstructed Abel density, the tracked contour, and the reconstructed blackening factor with their exact counterparts
already provides strong evidence that the conditions hold on the sampled
branches; we therefore do not display a separate set of Abel-reduction
validation plots.  When no benchmark is available, a practical \emph{a posteriori}
test is to substitute the reconstructed metric into the original width and
area integrals, using the same contour and square-root sheet.  Recovering
\(\Delta t(E)\) directly tests the Abel reduction, whereas recovering
\(S_{\rm reg}(\Delta t)\) tests the full forward entropy
construction.  These checks validate the complex-coordinate reduction on the
sampled branch, but they neither determine the saddle-selection rule nor establish existence or
uniqueness for the full metric inverse problem.

\acknowledgments
We would like to thank Ping Gao, Run-Qiu Yang, and Zhenkang Lu for helpful
discussions. SFW is supported by NSFC grants No.~12275166 and No.~12311540141.

\appendix
\section{Numerical Settings}
\label{app:numerical_parameters}

This appendix collects the numerical details of the forward and reconstruction
algorithms.  Tables~\ref{tab:forward_numerical_parameters} and
\ref{tab:reconstruction_numerical_parameters} summarize their principal
parameters.  We then specify the smoothing and repulsive forces, update rule,
and model-dependent settings used in the elastic-band relaxation of the
tracked complex contours.

\begin{table}[H]
    \centering
    \caption{Principal numerical parameters of the forward algorithm.}
    \label{tab:forward_numerical_parameters}
    \small
    \setlength{\tabcolsep}{4pt}
    \renewcommand{\arraystretch}{1.15}
    \begin{tabular}{@{}>{\raggedright\arraybackslash}p{0.27\textwidth}
                        >{\raggedright\arraybackslash}p{0.22\textwidth}
                        >{\raggedright\arraybackslash}p{0.22\textwidth}
                        >{\raggedright\arraybackslash}p{0.22\textwidth}@{}}
        \hline
        Parameter & SAdS$_4$ & RN-AdS$_4$ & Gubser--Rocha \\
        \hline
        Model parameter(s)
            & $z_h=1$
            & $Q=0.3,\,1.2,\,1.7$
            & \shortstack[l]{$\mu=0.5,\,1.0,\,1.5$} \\
        Initial turning point $z_t^{(0)}$
            & $0.01i$ & $0.02i$ & $0.05i$ \\
        Contour nodes
            & \shortstack[l]{initial $80$\\adaptive $200$--$1000$}
            & \shortstack[l]{initial $80$\\adaptive $200$--$1000$}
            & \shortstack[l]{initial $80$\\adaptive $200$--$1000$} \\
        Turning-point step size
            & $10^{-6}$--$0.02$
            & $10^{-6}$--$0.02$
            & $10^{-6}$--$0.02$ \\
        Maximum tracing steps
            & $500$ & $500$ & $350$ \\
        Turning-point cutoff $\epsilon_{\rm tp}$
            & $10^{-5}$ & $10^{-5}$ & $10^{-5}$ \\
        Quadrature absolute tolerance
            & $10^{-9}$--$10^{-12}$
            & $10^{-9}$--$10^{-12}$
            & $10^{-9}$--$10^{-12}$ \\
        \hline
    \end{tabular}
\end{table}

\begin{table}[H]
    \centering
    \caption{Principal numerical parameters of the reconstruction algorithm.}
    \label{tab:reconstruction_numerical_parameters}
    \small
    \setlength{\tabcolsep}{4pt}
    \renewcommand{\arraystretch}{1.15}
    \begin{tabular}{@{}>{\raggedright\arraybackslash}p{0.27\textwidth}
                        >{\raggedright\arraybackslash}p{0.22\textwidth}
                        >{\raggedright\arraybackslash}p{0.22\textwidth}
                        >{\raggedright\arraybackslash}p{0.22\textwidth}@{}}
        \hline
        Parameter & SAdS$_4$ & RN-AdS$_4$ & Gubser--Rocha \\
        \hline
        Reconstruction target
            & $\widehat H,\ z(u),\ f(z)$
            & $\widehat H,\ z(u),\ f(z)$
            & $H(W)$ only \\
        Model parameter(s)
            & $z_h=1$
            & $Q=1.2$
            & \shortstack[l]{$\mu=1.0$} \\
        Initial turning point $z_t^{(0)}$
            & $0.01i$ & $0.02i$ & $0.05i$ \\
        Turning-point step size
            & $10^{-6}$--$0.01$
            & $10^{-6}$--$0.02$
            & $10^{-6}$--$0.02$ \\
        Contour nodes
            & \shortstack[l]{initial $80$\\adaptive $400$--$2000$}
            & \shortstack[l]{initial $80$\\adaptive $400$--$2000$}
            & \shortstack[l]{initial $80$\\adaptive $200$--$1000$} \\
        Maximum tracing steps
            & $500$ & $800$ & $450$ \\
        Abel grid points
            & $1800$ & $1800$ & $3000$ \\
        Turning-point cutoff $\epsilon_{\rm tp}$
            & $3\times10^{-6}$ & $3\times10^{-6}$ & $10^{-5}$ \\
        Quadrature tolerance (absolute/relative)
            & $10^{-11}/10^{-11}$
            & $10^{-11}/10^{-11}$
            & $10^{-11}/10^{-11}$ \\
        Path interpolation
            & PCHIP & PCHIP & PCHIP \\
        Radial ODE solver
            & DOP853
            & DOP853
            & not used \\
        ODE tolerance (relative/absolute)
            & $10^{-10}/10^{-12}$
            & $10^{-10}/10^{-12}$
            & not used \\
        \hline
    \end{tabular}
\end{table}

\paragraph{Elastic-band relaxation.}
The discretized contour \(z_j\in\mathbb C\), with fixed endpoints, is relaxed using the smoothing and repulsive forces
\begin{equation}
\begin{aligned}
F_j^{\mathrm{sm}}
&=k_{\mathrm{sm}}(z_{j-1}-2z_j+z_{j+1}),\\
F_j^{\mathrm{rep}}
&=k_{\mathrm{rep}}\sum_\alpha
\Theta(R-r_{j\alpha})
\left(1-\frac{r_{j\alpha}}{R}\right)^2
\frac{z_j-\zeta_\alpha}{r_{j\alpha}^3},
\qquad
r_{j\alpha}=|z_j-\zeta_\alpha|,
\end{aligned}
\end{equation}
where \(\zeta_\alpha\) denotes the poles and branch points. The contour is updated as
\begin{equation}
z_j^{(n+1)}
=z_j^{(n)}+\eta
\left(F_j^{\mathrm{sm}}+F_j^{\mathrm{rep}}\right).
\end{equation}
The repulsive prefactor is capped at \(100\). For SAdS$_4$/RN--AdS$_4$, we use
\((k_{\mathrm{sm}},k_{\mathrm{rep}},R,\eta)
=(0.5,0.1,10^{-3},0.05)\),
with \(60\) forward and \(100\) reconstruction sweeps. For Gubser--Rocha, we use
\((0.1,0.1,0.04,0.05)\)
with \(150\) sweeps in both cases. Each relaxation is followed by adaptive resampling.

\section{Verification of the complex-coordinate Abel reduction}
\label{app:btz_abel_conditions}

This appendix checks the conditions of Abel reduction in
Sec.~\ref{subsec:abel_complex_coordinate} for the exact BTZ metric.

Consider the transformed variable
\begin{equation}
    W_{\rm C}(z)=\frac{1-z^2}{z^2}=z^{-2}-1.
\end{equation}
A real positive boundary time determines
\begin{equation}
    \Delta t>0
    \quad\Longrightarrow\quad
    E=-\coth\!\left(\frac{\Delta t}{2}\right)\in(-\infty,-1)
    \quad\Longrightarrow\quad
    x(E)\equiv -E^2\in(-\infty,-1).
\end{equation}
The turning-point condition \(W_{\rm C}(z_t)=x(E)\) then gives
\begin{equation}
    z_t^2=-\frac{1}{E^2-1}<0.
\end{equation}
Thus real \(\Delta t>0\) restricts the turning point to the imaginary axis.
Selecting the upper vacuum-connected branch fixes
\begin{equation}
    z_t=i y_t,
    \qquad
    y_t=\frac{1}{\sqrt{E^2-1}}>0.
\end{equation}
As \(\Delta t\) increases from \(0^+\) to \(+\infty\), \(E\) increases from
\(-\infty\) to \(-1\), \(x(E)\) increases from \(-\infty\) to \(-1\), and
\(z_t\) moves from \(0\) toward \(+i\infty\).

For each such turning point, the one-way contour may be taken directly from
the boundary to \(z_t\) along the positive imaginary axis,\footnote{On selecting the contour, we have taken the cutoff \(\epsilon\to0\) since the BTZ time-width integrand is regular in this limit.}
\[
    \bar C(0,z_t):\qquad z=iy,\qquad 0\leq y\leq y_t .
\]
Under the map \(W_{\rm C}(z)=z^{-2}-1\),
\begin{equation}
    W_{\rm C}(iy)=-1-y^{-2},
    \qquad 0<y\leq y_t.
\end{equation}
Consequently, as \(z=iy\) runs from the boundary \(y\to0^+\) to the turning
point \(y=y_t\), its image runs monotonically from \(W\to-\infty\) to
\(W=x(E)\). Hence \(E(\Delta t)\), \(z_t(\Delta t)\), and
\(x(E)=W_{\rm C}(z_t)\) are equivalent parameterizations of the same
one-parameter family. All the corresponding \(W\)-paths lie on
\((-\infty,-1)\), and varying \(\Delta t\) changes only their endpoint
\(x(E)\).

The width integrand has horizon poles at \(z=\pm1\) and square-root branch
points at \(z=\pm i y_t\), with \(+i y_t=z_t\) the allowed endpoint. The
distances from the chosen contour to the horizon poles and to the other branch
point are bounded below by
\begin{equation}
    d_{\rm pole}=1,
    \qquad
    d_{\rm br}=y_t>0.
\end{equation}
The contour therefore crosses no forbidden singularity, and the square-root
sheet remains fixed. Moreover,
\begin{equation}
    \frac{dW_{\rm C}}{dz}=-\frac{2}{z^3}\neq0
    \quad (0<|z|\leq y_t),
    \qquad
    z(W)=\frac{i}{\sqrt{-1-W}}.
\end{equation}
The inverse is single valued and differentiable at every finite point of the
path and satisfies \(z(W)\to i0^+\) as \(W\to-\infty\). The chosen contour
therefore provides the representative required in
Sec.~\ref{subsec:abel_complex_coordinate}, and the Abel-reduction conditions
hold for every finite \(\Delta t>0\).


 \bibliographystyle{JHEP}
 \bibliography{bib2.bib}

\end{document}